\documentclass[usenatbib,useAMS,usedcolumn]{mnras} 
\usepackage{times,epsfig,amsmath}

\hypersetup{pdfauthor={Ewan O'Sullivan},
            pdftitle={Molecular gas along the old radio jets of the cluster-central type~2 quasar IRAS~09104+4109},
            pdfkeywords={galaxies: clusters: individual (CL~09104+4109), galaxies: individual (IRAS~09104+4109),  galaxies: quasars: general,  galaxies: clusters: intracluster medium, X-rays: galaxies: clusters},
            bookmarksnumbered=true}
\pdfoutput=1

\newcommand{\arcm}{\hbox{$^\prime$}}

\newcommand{\degree}{\hbox{$^\circ$}}

\newcommand{\chandra}{\emph{Chandra}}

\newcommand{\hst}{\emph{HST}}
\newcommand{\arcs}{\mbox{\arcm\arcm}}

\newcommand{\Msol}{\ensuremath{\mathrm{~M_{\odot}}}}
\newcommand{\Msolpyr}{\ensuremath{\mathrm{~M_{\odot}~yr^{-1}}}}

\newcommand{\s}{\ensuremath{\mbox{~s}}}
\newcommand{\ps}{\ensuremath{\s^{-1}}}
\newcommand{\cm}{\ensuremath{\mbox{~cm}}}

\newcommand{\pcmcu}{\ensuremath{\cm^{-3}}}
\newcommand{\kev}{\ensuremath{\mbox{~keV}}}
\newcommand{\kevcmsq}{\ensuremath{\kev\cm^{2}}}
\newcommand{\km}{\ensuremath{\mbox{~km}}}
\newcommand{\Mpc}{\ensuremath{\mbox{~Mpc}}}
\newcommand{\pMpc}{\ensuremath{\Mpc^{-1}}}
\newcommand{\kmpspMpc}{\ensuremath{\km \ps \pMpc\,}}
\newcommand{\erg}{\ensuremath{\mbox{~erg}}}
\newcommand{\ergps}{\ensuremath{\erg \ps}}

\newcommand{\kmps}{\ensuremath{\km \ps}}

\newcommand{\gtsim}{\,\rlap{\raise 0.5ex\hbox{$>$}}{\lower 1.0ex\hbox{$\sim$}}\,}

\newcommand{\IRAS}{IRAS~09104+4109}

\begin{document}

\title[
CO in IRAS~09104+4109
]{
Molecular gas along the old radio jets of the cluster-central type~2 quasar IRAS~09104+4109 
}

\author[Ewan O'Sullivan et al.]
{Ewan O'Sullivan$^{1}$\thanks{eosullivan@cfa.harvard.edu},
  Fran\c{c}oise Combes$^{2,3}$,
  Arif Babul$^4$,
  Scott Chapman$^5$\newauthor
  Kedar A. Phadke$^6$,
  Gerrit Schellenberger$^1$,
  Philippe Salom\'{e}$^2$\\
$^1$ Center for Astrophysics $|$ Harvard \& Smithsonian, 60 Garden
  Street, Cambridge, MA 02138, USA \\
$^2$ LERMA, Observatoire de Paris, CNRS, PSL Univ., Sorbonne Univ., 75014 Paris, France\\
$^3$ Coll\`{e}ge de France, 11 place Marcelin Berthelot, 75005 Paris, France\\
$^4$ Department of Physics and Astronomy, University of Victoria, Victoria, BC, V8W 2Y2, Canada\\
$^5$ Department of Physics and Atmospheric Science, Dalhousie University, Halifax, NS B3H 3J5, Canada\\
$^6$ Department of Astronomy, University of Illinois, 1002 West Green Street, Urbana, IL 61801, USA\\
}

\date{Accepted 2021 September 28; Received 2021 September 24; in original form 2021 July 8}

\pagerange{\pageref{firstpage}--\pageref{lastpage}} \pubyear{2020}

\label{firstpage}

\maketitle

\begin{abstract}
We present Northern Extended Millimeter Array (NOEMA) CO(2-1) maps of the $z$=0.4418 cluster-central QSO IRAS~09104+4109, which trace $\sim$4.5$\times$10$^{10}$\Msol\ of molecular gas in and around the galaxy. As in many low redshift cool core clusters, the molecular gas is located in a series of clumps extending along the old radio jets and lobes. It has a relatively low velocity dispersion (336$^{+39}_{-35}$\kmps\ FWHM) and shows no velocity gradients indicative of outflow or infall. Roughly half the gas is located in a central clump on the northeast side of the galaxy, overlapping a bright ionized gas filament and a spur of excess X-ray emission, suggesting that this is a location of rapid cooling. The molecular gas is unusually extended, out to $\sim$55~kpc radius, comparable to the scale of the filamentary nebula in the Perseus cluster, despite the much higher redshift of this system. The extent falls within the thermal instability radius of the intracluster medium (ICM), with t$_{\rm cool}$/t$_{\rm ff}$$<$25 and t$_{\rm cool}$/t$_{\rm eddy}$$\sim$1 within $\sim$70~kpc. Continuum measurements at 159.9~GHz from NOEMA and 850~$\mu$m from the JCMT SCUBA-2 show excess far infrared emission, which we interpret as free-free emission arising from the ongoing starburst. These observations suggest that ICM cooling is not strongly affected by the buried QSO, and that cooling from the ICM can build gas reservoirs sufficient to fuel quasar-mode activity and drive the reorientation of the central AGN.
\end{abstract}

\begin{keywords}
galaxies: clusters: individual (CL~09104+4109) --- galaxies: individual (IRAS~09104+4109) --- galaxies: quasars: general --- galaxies: clusters: intracluster medium --- X-rays: galaxies:clusters
\end{keywords}

\section{Introduction}
\label{sec:intro}

It is now well established that heating by active galactic nuclei (AGN) plays the dominant role in balancing radiative cooling in the cores of relaxed galaxy clusters. In the nearby Universe, AGN in group- and cluster-dominant galaxies are almost always radio-mode systems, with numerous X-ray observations convincingly showing the connection between the radio jets and cavities inflated in the hot intra-cluster medium \citep[ICM, see e.g.,][]{McNamaraNulsen07,Gittietal12,Fabian12}. The brightest cluster galaxies (BCGs) of cool core clusters are often surrounded by extended filamentary optical emission-line nebulae \citep[e.g.,][]{Fabianetal03,Crawfordetal05,Crawfordetal05b,McDonaldetal10,McDonaldetal11,Lakhchauraetal18}, in many cases spatially correlated with molecular gas \citep[e.g.,][]{Salomeetal06,Salomeetal11,Russelletal16,Vantyghemetal16,Russelletal17a,Russelletal17b,Vantyghemetal17,Vantyghemetal18,Olivaresetal19}, soft X-ray structures \citep[e.g.,][]{Werneretal14} and young stars \citep[e.g.,][]{Canningetal14}. These filamentary nebulae typically extend a few tens of kiloparsecs and can contain 10$^9$-10$^{11}$\Msol\ of molecular gas. 

The ionized and molecular gas is considered to be the product of cooling from the ICM, and the most likely fuel source for the AGN of the BCG. The nebulae tend to be found in systems where the central cooling time is short ($<$1~Gyr) and the ICM entropy is low \citep[$<$30~keV~cm$^2$,][]{Cavagnoloetal08,Raffertyetal08}, and there is evidence that AGN jet power is correlated with the molecular gas mass \citep{Babyketal19} and, tentatively, with the H$\alpha$ emission from the ionized component \citep{Lakhchauraetal18}. These correlations have been interpreted as indicating a threshold beyond which the ICM gas becomes thermally unstable and can potentially condense out \citep{Sharmaetal12,McCourtetal12,Gasparietal13,LiBryan14a,LiBryan14b,Prasadetal15}.

However, there is still some debate as to whether the cooled material simply forms \textit{in situ}, or whether some other factor is needed to trigger the condensation \citep{McNamaraetal16,Prasadetal17,Prasadetal18,Voitetal17,Hoganetal17,Gasparietal18,Qiuetal20}. The filaments of cooled gas are often found around the ICM cavities associated with radio lobes, or at smaller radii, in some cases lying along the presumed path of the lobes rise. This raises the possibility that as the lobes inflate and rise away from the AGN, they uplift cool ICM gas from the core \citep{Revazetal08,Popeetal10} which is either already thermally unstable or becomes so as it rises, and thus condenses out along the line of ascent \citep{McNamaraetal16,Qiuetal20}. Uplift of ionized gas already present in the core may also be possible. Uplift of molecular gas would be more difficult given its high density, unless it is strongly connected to the ICM via surrounding layers of ionized gas \citep{Lietal18} or magnetic fields \citep{McCourtetal15}. Another possibility is that thermally unstable gas at larger radii is disturbed by the expansion of jets or lobes, with the induced turbulence and compression of the ICM triggering condensation \citep{Gasparietal11, Gasparietal12,Lietal15,Prasadetal15,Prasadetal17}.

While almost all nearby cluster-central AGN are low accretion rate, radiatively inefficient radio galaxies, at higher redshift many AGN are radiatively efficient quasar-type systems \citep[e.g.,][]{HlavacekLarrondoetal13} raising the question of how radiative cooling was regulated in clusters at earlier times. Unfortunately, very few low-redshift examples of cluster-central quasars are known. The BCG of the Phoenix Cluster hosts both a quasi-stellar object (QSO) and an extended filamentary nebula wrapped around two radio lobes \citep{McDonaldetal19}. The nebula contains $\sim$2$\times$10$^{10}$\Msol\ of molecular gas, fuelling star formation at the rate of 500-800\Msolpyr, but the AGN accretion rate is thought to be falling, causing a shift from quasar to jet-mode activity \citep{Russelletal17a,Prasadetal20}. A second cluster, H1812+643, hosts a type~I (unobscured) quasar whose photon flux may be causing Compton cooling in the ICM \citep{Russelletal10,Reynoldsetal14,Walkeretal14}. However, its fuel source is less clear; while the cooling flow should provide a plentiful supply of cooled ICM gas, there are conflicting reports of its molecular gas content \citep{Combesetal11,Aravenaetal11}.

A third cluster-central quasar system is \IRAS, a Hyper-Luminous Infrared Galaxy (HyLIRG) located at the centre of the cluster CL~09104+4109 (also known as MACS~J0913.7+4056). As well as hosting what may be the most luminous Compton-thick (type~2, obscured) QSO at $z<0.5$ \citep{Farrahetal16}, the BCG has $\sim$70~kpc FR~I radio jets whose radio spectrum suggests they were formed in a period of activity which ceased 120-160~Myr ago and a Very Long Baseline Array (VLBA) observation reveals a 200~pc scale inner radio double indicative of a more recent outburst, in the BCG nucleus \citep[hereafter OS12]{OSullivanetal12special}. The QSO is obscured, but polarized optical emission traces its ionization cones, which are misaligned from the radio jets \citep{Hinesetal99} indicating a change in orientation. IRAM~30m CO(2-1) observations suggest the BCG contains 3.2$\times$10$^9$\Msol\ of molecular gas, but only a relatively small amount of dust \citep[$<$5$\times$10$^7$\Msol,][]{Combesetal11}. The BCG appears to have undergone a burst of star formation 70-200~Myr ago \citep{Bildfelletal08,Pipinoetal09} with a second burst, starting $<$50~Myr ago, producing a current star formation rate of 110$^{+35}_{-28}$\Msolpyr\ \citep{Farrahetal16}. \textit{Hubble Space Telescope} (\hst) imaging reveals a number of ionized gas filaments around the BCG. Ground-based Integral Field Unit (IFU) observations of the brightest of these, a plume of [O\textsc{iii}] emission extending $\sim$27~kpc from the BCG, show that its velocity is only $\sim$100\kmps\ offset from that of the galaxy \citep{CrawfordVanderriest96}. Such small velocity offsets are typical of the filamentary nebulae seen in nearby cool-core clusters.

In this paper we describe new Northern Extended Millimeter Array (NOEMA) and James Clerk Maxwell Telescope (JCMT) observations of \IRAS, designed to probe the molecular gas and dust environment of the galaxy and determine where, and how much, gas is cooling out of the ICM to fuel the quasar. Throughout the paper we assume a flat cosmology with H$_0$=73\kmpspMpc, $\Omega_\Lambda$=0.73 and $\Omega_{\rm M}$=0.27. We adopt a redshift of $z$=0.4418 for the galaxy, which gives an angular scale of 1\arcs=5.5~kpc, luminosity distance D$_{\rm L}$=2372~Mpc and angular distance D$_{\rm A}$=1140~Mpc, for ease of comparison with OS12.

\section{Observations and Data Reduction}
\label{sec:obs}

\subsection{NOEMA}
\IRAS\ was observed by NOEMA on 2018 Nov 29, 2019 Dec 29 and 2019 Jan 7 as
program W18DB (P.I. O'Sullivan). All ten dishes were used, with the array in C configuration. We set the phase center of the observations equal to the target coordinates, R.A.~=~09$^h$:13$^m$:45\farcs 489 and Dec.~=~40\degree :56\arcm :28\farcs 22. We also adopted a tuning frequency of 159.9~GHz, which corresponds to the observer frame frequency of the redshifted CO(2-1) line, given the QSO
redshift.

The baseline range was 24--368~m. The program was executed in good to typical winter weather conditions, with system temperature $T_{\rm sys}=100-220$~K  and a precipitable water vapour column between 2~mm and 7~mm. We used the PolyFix correlator which covers a total bandwidth of 15.5~GHz split between the lower and upper side band, covering the ranges between 142.5-150.2~GHz and 157.8-165.5~GHz, respectively. The sources 0923+392, 0906+432 and 0916+390 were used as phase and amplitude calibrators, while 3C84 and 3C279 served as bandwidth and flux calibrators. We reduced the data using the \textsc{clic} package of the \textsc{gildas} software \citep{Pety05,GILDAS13}. The final ($u$,$v$) tables correspond to 6 h of total integration time, of which 4~h were on source.

We imaged the visibilities using the \textsc{mapping} software from \textsc{gildas}. At the tuning frequency of 159.9~GHz, the half power primary beam width is 31.5\arcs. We adopted natural weighting, yielding a synthesized beam of 1.55\arcs$\times$1.14\arcs\ with PA=-16.7\degree. We then re-binned the spectral axis at a resolution of 30\kmps\ ($\sim16$~MHz at 159.9 GHz). The resulting rms is 0.72~mJy~beam$^{-1}$.

\subsubsection{NOEMA positional accuracy}
The astrometry of the NOEMA data is expected to be highly accurate, but as a precaution, we confirmed it by imaging the continuum emission and comparing its centroid position with imaging from other wavelengths. OS12 present images of the system in a variety of bands, including \textit{HST} and Canada-France-Hawaii Telescope optical data, \textit{Chandra} X-ray images, Giant Metrewave Radio Telescope (GMRT) and Very Large Array (VLA) data in a total of six bands covering 240~MHz to 4.8~GHz, and a 1.4~GHz VLBA image. The 12~hr VLBA observation (P.I. Wrobel, project code BH0110) was phase-referenced, performed in the nodding style \citep{Wrobeletal00}, switching between alternate $\sim$180~s scans of the target and $\sim$120~s scans of the phase calibrator. It can therefore be expected to have milliarcsecond astrometric accuracy. The VLBA position was used for the phase centre of the NOEMA observation, and the position of the NOEMA continuum source agrees well with the optical centroid of the galaxy, the central component of the megahertz and gigahertz radio emission, the hard X-ray source associated with the QSO, and the VLBA position. We are therefore confident in the astrometry of the NOEMA observation, and of the supporting data.

\subsection{JCMT}
\IRAS\ was observed with the Submillimeter Common-User Bolometer Array 2 (SCUBA-2) for a total of 6 hours on source, in three programs from 2012 through 2018, the majority in our own program M15BL114 (P.I. Chapman): M18BP056 --- 30~min, M15BL114 --- 5~hr, M12AC15 --- 30~min. Observations were conducted in Band 1-2 weather conditions ($\tau_{\rm 225~GHz}$$\sim$0.04-0.08). The mapping centre of the SCUBA-2 field was set to the same coordinates as the NOEMA phase centre. A standard 3\arcm\ diameter \textsc{daisy} mapping pattern was used \citep[e.g.,][]{Kackleyetal10} which keeps the centre on one of the four SCUBA-2 sub-arrays at all times during the exposure.

The twelve individual 30~min scans were reduced using the dynamic iterative map-maker of the \textsc{smurf} package \citep{Jennessetal13,Chapinetal13}. Maps from independent scans were co-added in an optimal stack using the variance of the data contributing to each pixel to weight spatially aligned pixels. Finally, since we are interested in (generally faint) extragalactic point sources, we applied a beam matched filter to improve point source detectability, resulting in a map that is convolved with an estimate of the 850 or 450~$\mu$m beam.

The sky opacity at JCMT has been obtained by fitting extinction models to hundreds of standard calibrators observed since the commissioning of SCUBA-2 \citep{Dempseyetal12}. These maps have been converted from pW to Jy using the standard conversion factors (FCFs) of FCF$_{450}$=491~Jy~beam$^{-1}$~pW$^{-1}$ and FCF$_{850}$=547~Jy~beam$^{-1}$~pW$^{-1}$, with effective 450 and 850~$\mu$m beam sizes of 10\arcs and 15\arcs\ respectively \citep{Dempseyetal13}.

The variance map was derived for each pixel from the data time series \citep[as in, e.g.,][]{Koprowski15}. The RMS within the central 5\arcm\ diameter region varies from 0.45-0.55~mJy~beam$^{-1}$ at 850~$\mu$m and 3.4-4.1~mJy~beam$^{-1}$ at 450~$\mu$m. Our depths reached at both 850~$\mu$m and 450~$\mu$m and smaller beam sizes allow us to probe sources in the cluster core more effectively than the confusion limited \textit{Herschel} maps.

\section{Results}
\label{sec:res}

\subsection{CO line emission}
\label{sec:COres}

\begin{figure*}
\includegraphics[width=\columnwidth,viewport=50 150 570 660]{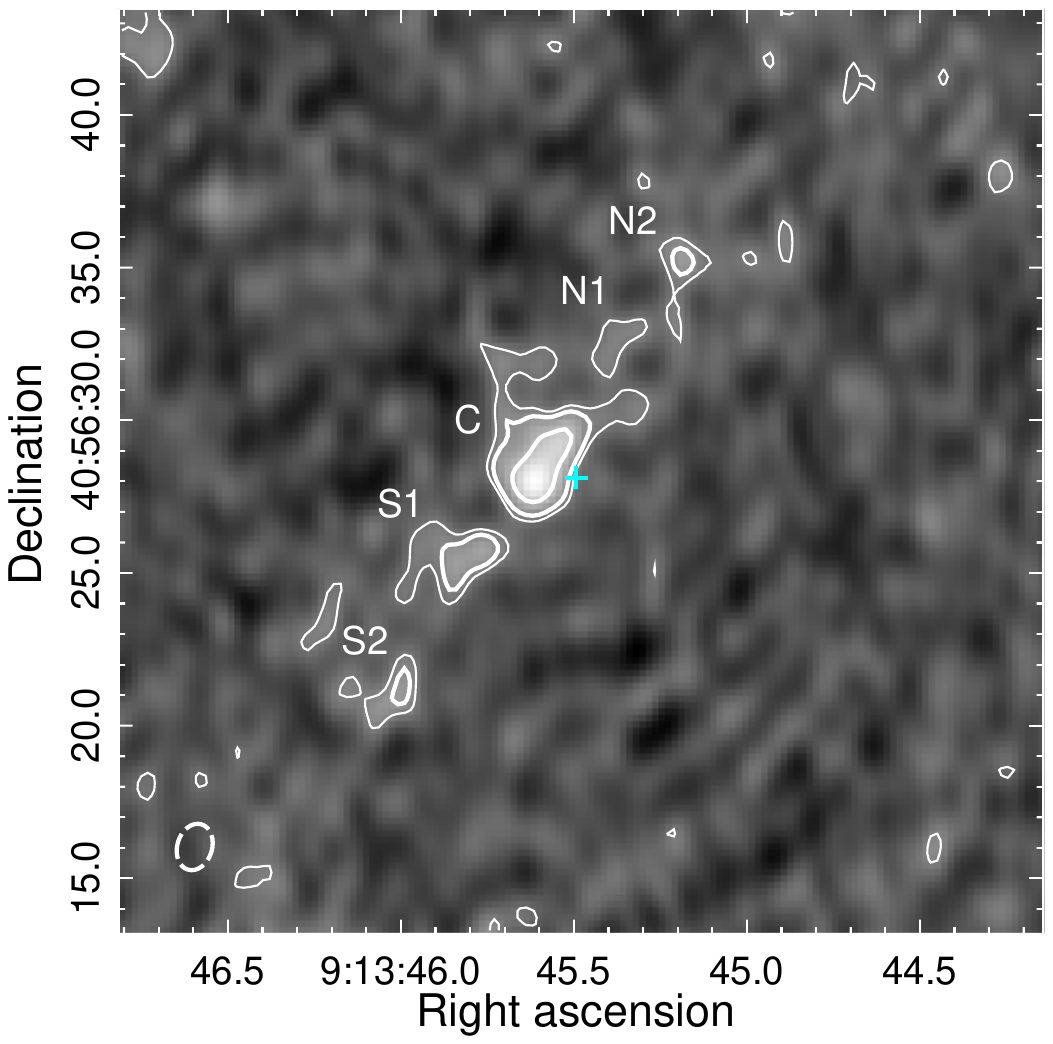}
\includegraphics[width=\columnwidth,viewport=50 150 570 660]{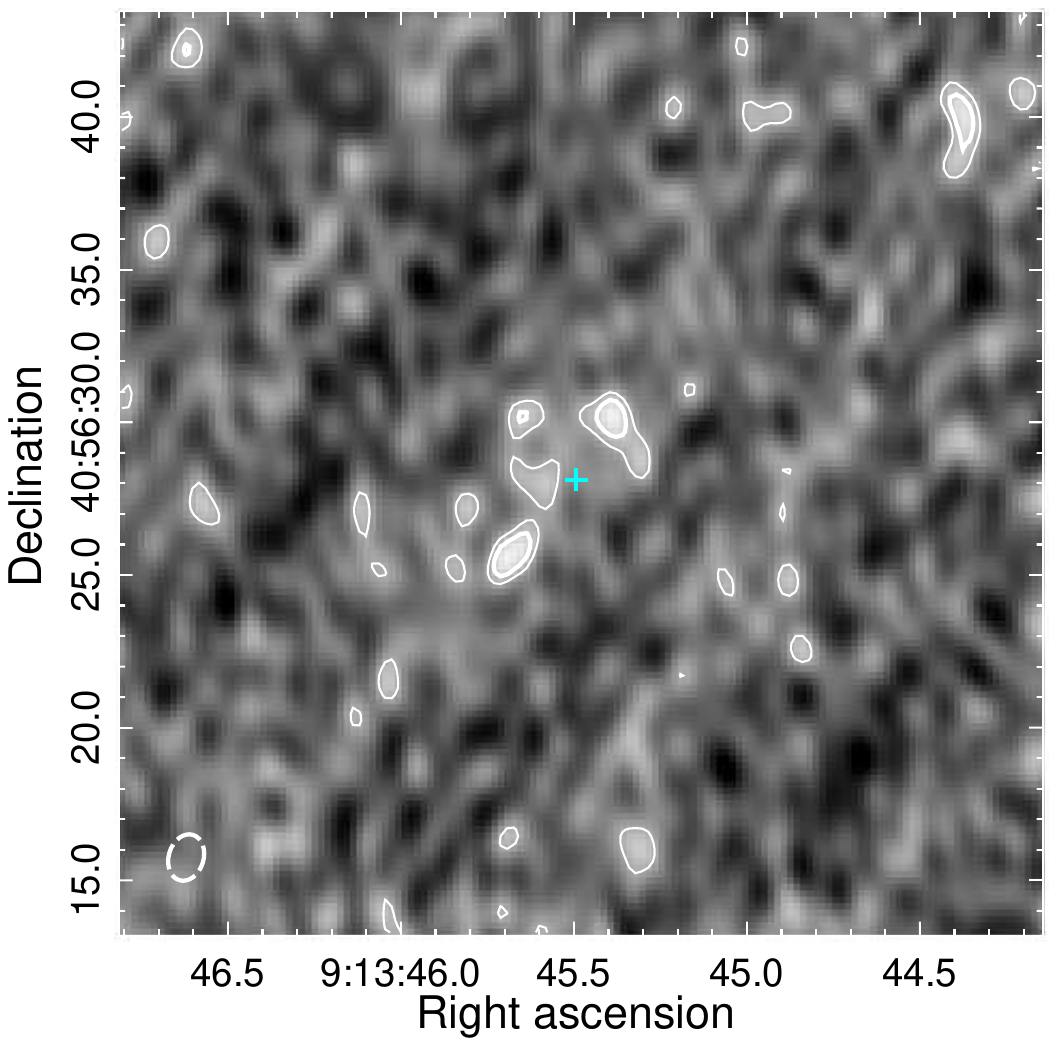}
\caption{\label{fig:COflux}NOEMA moment 0 maps of integrated CO(2-1) flux, within -300 to 210\kmps\ of the systemic velocity (\textit{left}) and at -1440 to -1470\kmps\ (\textit{right}). Contours mark levels 2, 3 and 6$\sigma$ above the rms noise level (71.5~mJy~bm$^{-1}$\kmps\ for the left panel, 24.0~mJy~bm$^{-1}$\kmps\ for the right). The dashed ellipse indicates the NOEMA beam, and the cross the VLBA-determined position of the AGN. Note that the images have not been corrected for the NOEMA primary beam (HPBW=31.5\arcs) so sensitivity declines toward the edge of the field.}
\end{figure*}

An initial examination of the continuum-subtracted NOEMA data cube suggests that most of the CO(2-1) flux is found within a few hundred \kmps\ of the BCG. A number of emission clumps are visible in this velocity range, close to the BCG or the radio jets.

When analysing data cubes containing spectral line emission, it is common to ``clip'' the data, removing spectral channels where the signal falls below some significance threshold. As discussed in \citet{Dame11}, while this reduces the noise level of the data, it also introduces biases, most notably a negative bias owing to the removal of spectral line wings and low intensity extended emission. In many cases such a bias is acceptable, as it results in cleaner maps with well defined, highly-significantly detected clouds. However, for fainter targets, excessive clipping can result in the loss of scientifically meaningful structures. \citet{Dame11} advocate an alternate approach in which the data are smoothed spatially and spectrally, and a clipped version of this smoothed cube is used as a mask. The smoothing suppresses random noise peaks, while the significance of genuine emission, which can be expected to be in contiguous structures extending across multiple beams and/or spectral channels, should be enhanced. In creating images of our source, we experimented with this filtering method and with the choice of velocity range from which to extract data. We found that the best results were achieved using a mask Gaussian smoothed over 2 spectral channels and to $\sim$3\arcs\ spatial resolution, which was then clipped to remove any region with $<$2$\sigma$ significant emission in an individual channel. Images were extracted in the velocity range -300 to 210\kmps. In the resulting image, several clumps of emission are detected at $>$3$\sigma$ significance, and we note that these are also detected at this significance level in images made with the unfiltered data.

Figure~\ref{fig:COflux} shows the resulting image, at the full spatial resolution of the observation (maps of the individual masked channels can be found in Appendix~\ref{sec:chanmaps}). The emission is dominated by a large clump near the middle of the image (clump C), whose centre is detected at $>$6$\sigma$ significance. Three smaller clumps (S1-2, N2) are detected at $>$3$\sigma$ significance along a line extending roughly northwest-southeast. Note that the images have not been corrected for the NOEMA primary beam (HPBW=31.5\arcs) so  sensitivity declines toward the edge of the field of view. Contours mark the 2, 3 and 6$\sigma$ significance levels; we have included the 2$\sigma$ contour to show the hints of more extended emission around the various clumps, and additional emission between them, but we note that we we only consider 3$\sigma$ significant structures as detected in our analysis. Where 2$\sigma$ significant features are discussed, we emphasize that we cannot be sure that these are not noise features.

There is also a tentative detection of flux at $\sim$-1450\kmps. Figure~\ref{fig:COflux}~right shows the flux map for the combination of two spectral channels, -1440 and -1470\kmps. Two small clumps of emission are detected at $>$3$\sigma$ significance, positioned 2.5-3.5\arcs\ (14-19~kpc) northwest and southeast of the BCG centroid. As expected for such a narrow velocity range, the image is relatively noisy, with numerous small clumps at $>$2$\sigma$ significance, and one other 3$\sigma$ significant region larger than the beam size, in the northwest corner of the image. Based on the noise level and image size, we expect 1-2 false 3$\sigma$ detections in the field.

\begin{figure}
\includegraphics[width=\columnwidth,bb=45 220 560 730]{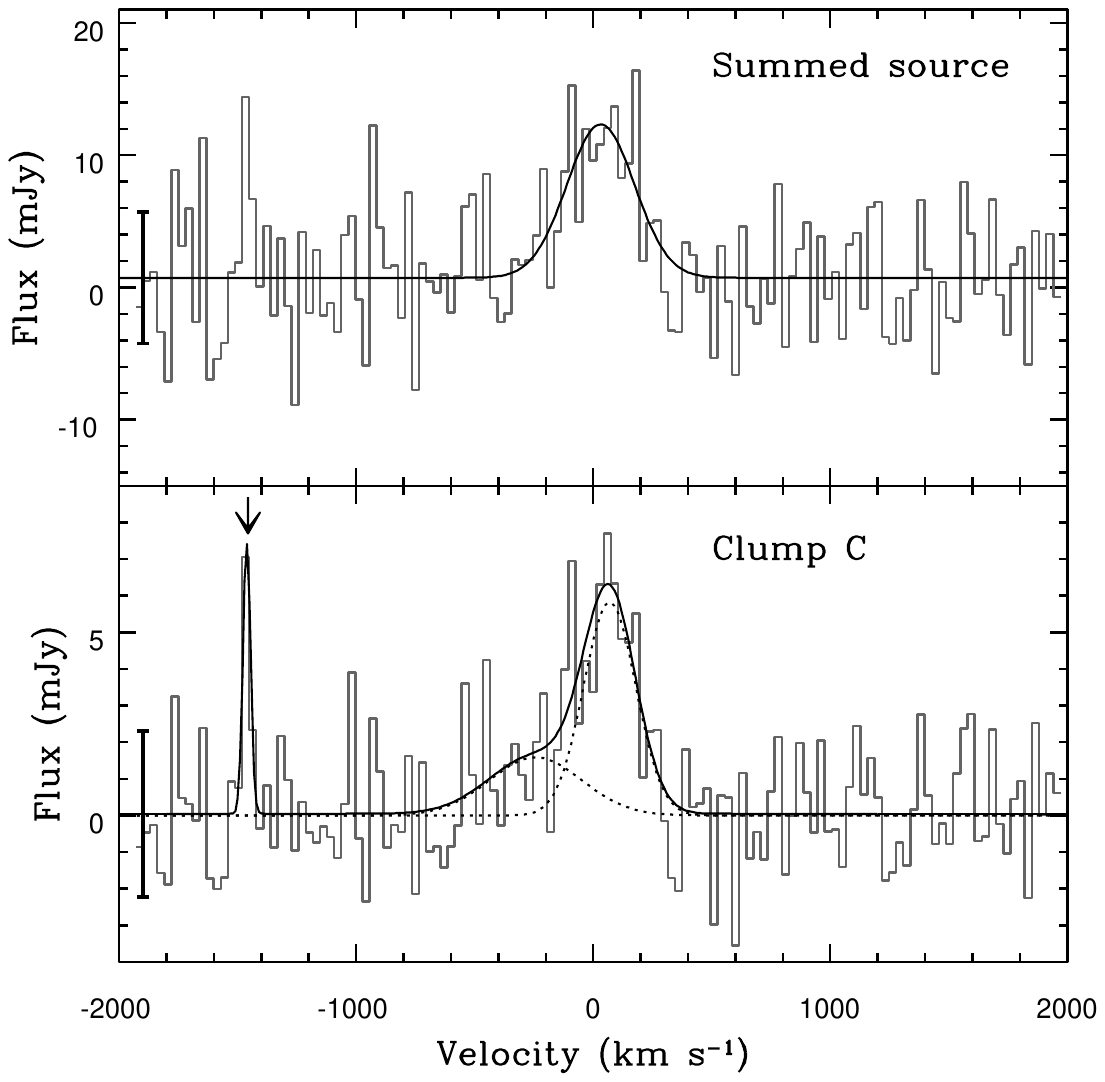}
\caption{\label{fig:spec}NOEMA spectra for the source as a whole (\textit{upper panel}, 4.97~mJy rms) and for clump~C (\textit{lower panel}, 2.27~mJy rms) shown as grey histograms, with best-fitting models overlaid in solid black. For clump~C, the two Gaussian components used to model the main peak are shown individually as dotted lines. An arrow marks the position of the high-velocity component. Error bars at -1950\kmps\ indicate the rms noise level of each spectrum.}
\end{figure}

We extracted spectra for the four clumps detected at $>$3$\sigma$ significance around the systemic velocity (C, N2, S1 \& S2), for a region enclosing all of those clumps, and for a region enclosing the high-velocity clumps close to the BCG. Figure~\ref{fig:spec} shows example spectra for clump~C and for the large region enclosing the source as a whole. For the summed source, the spectrum is noisy, but a central peak is clear, and this is reasonably well approximated by a Gaussian centred close to the systemic velocity (+31$\pm$17\kmps). The best-fitting model parameters are shown in Table~\ref{tab:fluxes}. For clump~C, the spectrum is more complex; the central peak has a tail extending to negative velocities, and a possible secondary spike corresponding to the high-velocity clumps is visible. We were able to model the spectrum using two Gaussian components for the main peak, and a narrow third Gaussian for the high-velocity component. With a small constant component to account for any remaining continuum flux (statistically consistent with zero), this provides a good fit to the data ($\chi^2$=129.18 for 120 degrees of freedom). Imaging flux in the velocity range -750 to -90\kmps\ suggests that the material forming the negative velocity tail is located primarily in the brightest peak of clump~C, with some possible extension to the southeast toward clump~S1. The material in the positive velocity peak is more widespread. The smaller clumps were all fitted using single Gaussians. Fit parameters for each clump are again shown in Table~\ref{tab:fluxes}.

\begin{table}
\caption{\label{tab:fluxes}Best fitting spectral model parameters and fluxes for the summed source and the component clumps.}
\begin{center}
\begin{tabular}{lcccc}
\hline
Component & \multicolumn{3}{c}{Gaussian} & M$_{\rm mol}$\\
 & $\langle$v$\rangle$ & FWHM & S$_{\rm CO}\Delta$v & \\
 & (\kmps) & (\kmps) & (Jy~\kmps) & (10$^{10}$\Msol)\\
\hline\\[-3mm]
summed & 31$\pm$17 & 336$^{+39}_{-35}$ & 4.15$^{+0.48}_{-0.45}$ & 4.56$^{+0.53}_{-0.49}$ \\
C & 70$^{+32}_{-42}$ & 259$^{+92}_{-77}$ & 1.61$^{+0.62}_{-0.75}$ & 1.77$^{+0.82}_{-0.68}$\\ 
  & -242$^{+178}_{-250}$ & 459$^{+198}_{-339}$ & 0.77$^{+0.69}_{-0.57}$ & 0.85$^{+0.76}_{-0.63}$\\ 
N2 & 44$^{+108}_{-97}$ & 320$^{+174}_{-119}$ & 0.15$^{+0.09}_{-0.08}$ & 0.16$^{+0.10}_{-0.08}$ \\ 
S1 & 2$\pm$42 & 315$^{+86}_{-68}$ & 0.58$^{+0.16}_{-0.15}$ & 0.64$^{+0.18}_{-0.16}$ \\ 
S2 & 8$^{+64}_{-62}$ & 278$^{+119}_{-140}$ & 0.27$\pm$0.12 & 0.30$\pm$0.13 \\ 
HV & -1456$^{+4}_{-5}$ & 40$^{+14}_{-31}$ & 0.50$^{+0.11}_{-0.09}$ & 0.55$^{+0.12}_{-0.10}$ \\ 
\hline
\end{tabular}
\end{center}
\end{table}

With the exception of clump C, all clumps have mean velocities consistent with the systemic velocity, and there is no clear trend in velocity along the line of clumps. The integrated flux in the spectrum for the source as a whole is marginally higher than the sum of the fluxes in the clumps, but consistent within the uncertainties. Based on the integrated flux measured from these models, we estimate the molecular gas mass (M$_{\rm mol}$) in the source as a whole, and in each clump, following \citet{OSullivanetal18} and adopting the conversion factor for nearby quiescent galaxies, $\alpha_{\rm CO}$=4.6 \citep{SolomonVandenbout05}. This suggests a total mass of molecular gas of $\sim$4.5$\times$10$^{10}$\Msol\ at the velocity of the BCG. However, we note that in ultra-luminous infrared galaxies (ULIRGs) and intensely star-forming systems CO is expected to be more luminous than in quiescent galaxies, owing to higher densities and temperatures in the molecular clouds. These conditions may hold in \IRAS, particularly in clump~C. \citet{Combesetal11} assumed $\alpha_{\rm CO}$=0.8 for their molecular gas mass estimate of this system. Similarly, for their estimate of the CO mass in and around the BCG of the Phoenix Cluster, \citet{Russelletal17} assumed a lower conversion factor based on the work of \citet{DownesSolomon98}. If we adopted the $\alpha_{\rm CO}$=0.8, we would find a total molecular gas mass of $\sim$7.9$\times$10$^{9}$\Msol, a factor 2.5$\times$ greater than the IRAM~30m mass estimate, but we note that we cannot be certain of the conditions in the gas with the available data.

\begin{figure}
\includegraphics[width=\columnwidth,bb=60 450 560 730]{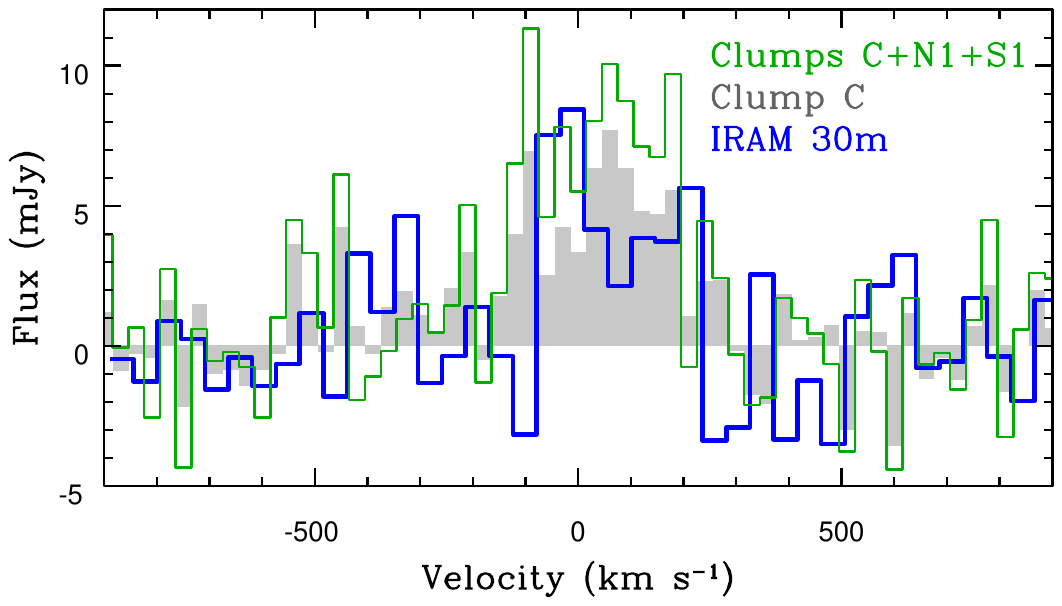}
\caption{\label{fig:IRAMspec}IRAM~30m CO(2-1) spectrum (heavy blue line) plotted over the NOEMA CO(2-1) spectrum for clump~C (shaded grey) and the summed spectrum of clumps C, N1 and S1 (thinner green line).}
\end{figure}

\subsubsection{Comparison with IRAM~30m detection}

Figure~\ref{fig:IRAMspec} shows a comparison of the IRAM~30m spectrum for \IRAS\ \citep[at slightly higher resolution that that presented by][]{Combesetal11} with the NOEMA spectrum of clump~C and the summed spectrum of clumps C+N1+S1. Note that clump N1 is only 2$\sigma$ significant in our images, but as it falls well within the IRAM~30m beam, we have included it to make sure any flux contribution from this region is included in the comparison. The IRAM~30m spectrum appears to be double-peaked, but when fitted with a single Gaussian is found to be relatively narrow (FWHM=225$\pm$55\kmps) with a flux of 1.4$\pm$0.4~Jy\kmps. The primary beam of the 30m telescope has a HPBW of 15.4\arcs\ at 159.9~GHz, roughly half the diameter of the NOEMA primary beam, so the older observation will not have included the outer clumps of emission we observe. Comparing the spectra, the overall line width in the IRAM~30m spectrum is similar to that of clump~C, or clumps~C+N1+S1, but the line height is more similar to that of clump~C alone. Comparing the Gaussian fits to the lines, the flux and line width from the IRAM~30m are similar to those of the dominant component of clump~C (FWHM=259$^{+92}_{-77}$\kmps, 1.61$^{+0.62}_{-0.75}$~Jy\kmps) and the mean velocities are comparable within the uncertainties ($\langle v\rangle_{\rm IRAM}$=32$\pm$30\kmps\ compared to $\langle v \rangle_{\rm clump  C}$=70$^{+32}_{-42}$\kmps). It seems possible, therefore, that the shallower IRAM 30m observation was dominated by emission from clump~C. For comparison the total flux in clumps C+N1+S1 is $\sim$3.28~Jy\kmps. Alternatively, the baseline continuum level of the IRAM~30m data may have been overestimated, leading to the narrower line width and CO(2-1) flux underestimate. In either case, it seems clear that the NOEMA interferometer observations are recovering all of the flux detected by the IRAM~30m and more.

\begin{figure*}
\includegraphics[width=0.4\textwidth,viewport=36 70 577 730]{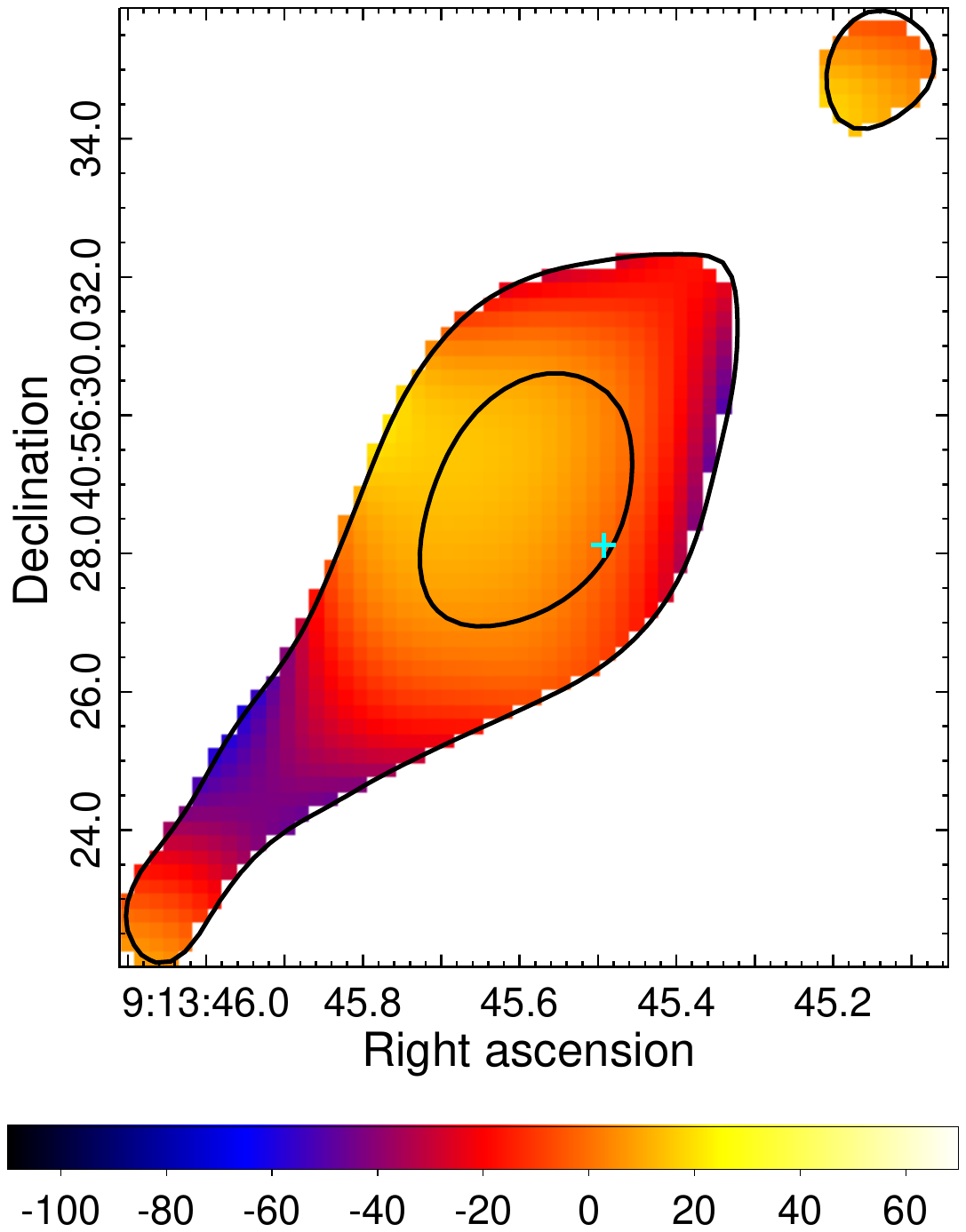}
\hspace{1cm}
\includegraphics[width=0.4\textwidth,viewport=36 70 577 730]{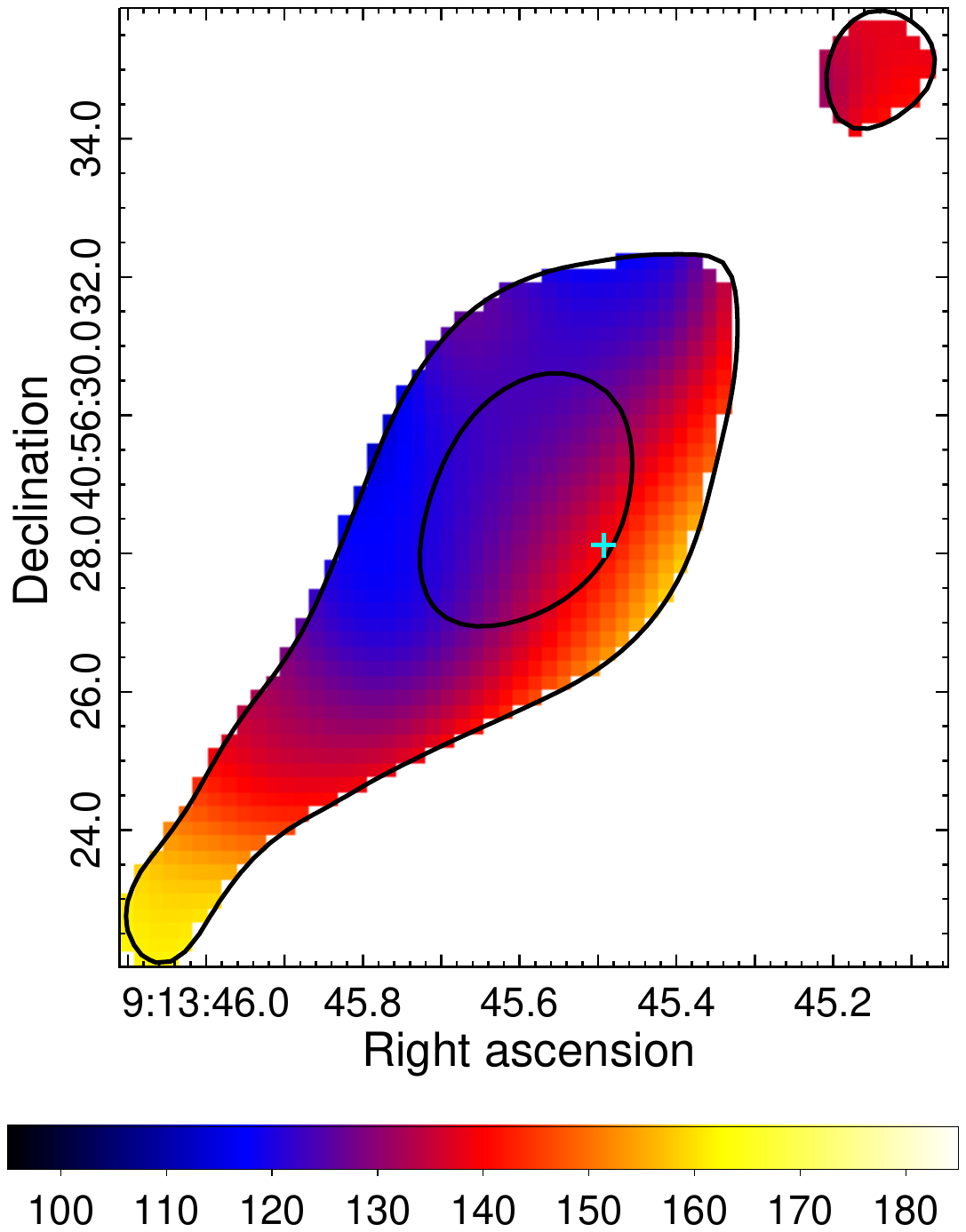}
\caption{\label{fig:COvel} Maps of smoothed NOEMA CO(2-1) moment 1 (velocity relative to the BCG, in \kmps) and 2 (velocity dispersion, in \kmps), made using a 3\arcs\ HPBW restoring beam. Both panels show the same field of view. Contours indicate flux 4 and 8$\sigma$ above the smoothed rms noise level. The cross indicates the VLBA-determined AGN position.}
\end{figure*}

\subsubsection{CO velocity maps}

At full resolution, maps of moments 1 and 2 (radial velocity and velocity dispersion) are very noisy, owing to the limited signal-to-noise of the data in the individual clumps. We therefore created images from a broader velocity range (-420 to 360\kmps) smoothed with a 3\arcs\ HPBW restoring beam, and applying the 2$\sigma$ clipping mask described above. We also reject all pixels whose flux level in the final moment 0 map is less than 4$\sigma$ significant. These images are shown in Figure~\ref{fig:COvel}. At this resolution, a general northwest-southeast structure is visible, with the central clump (which now includes its immediate neighbours) detected at high ($>$8$\sigma$) significance. A small clump is found to the northwest, corresponding to clump N2 in the full resolution image, while a tail extends to the southeast, its tip corresponding roughly with clump S2. The moment 1 map shows the relative velocity of the CO emission to be small compared to the BCG, generally $<$100\kmps. The moment 2 map shows generally small velocity dispersion in the gas, $\sim$110-130\kmps\, except on the western side of the main clump and in the southeast tail, where values $\sim$160\kmps\ are found. In the region overlapping the BCG centroid it is possibile that the higher velocity dispersions are associated with the gas in the deepest parts of the potential well of the BCG.

\begin{figure}
\includegraphics[width=\columnwidth,viewport=60 150 555 645]{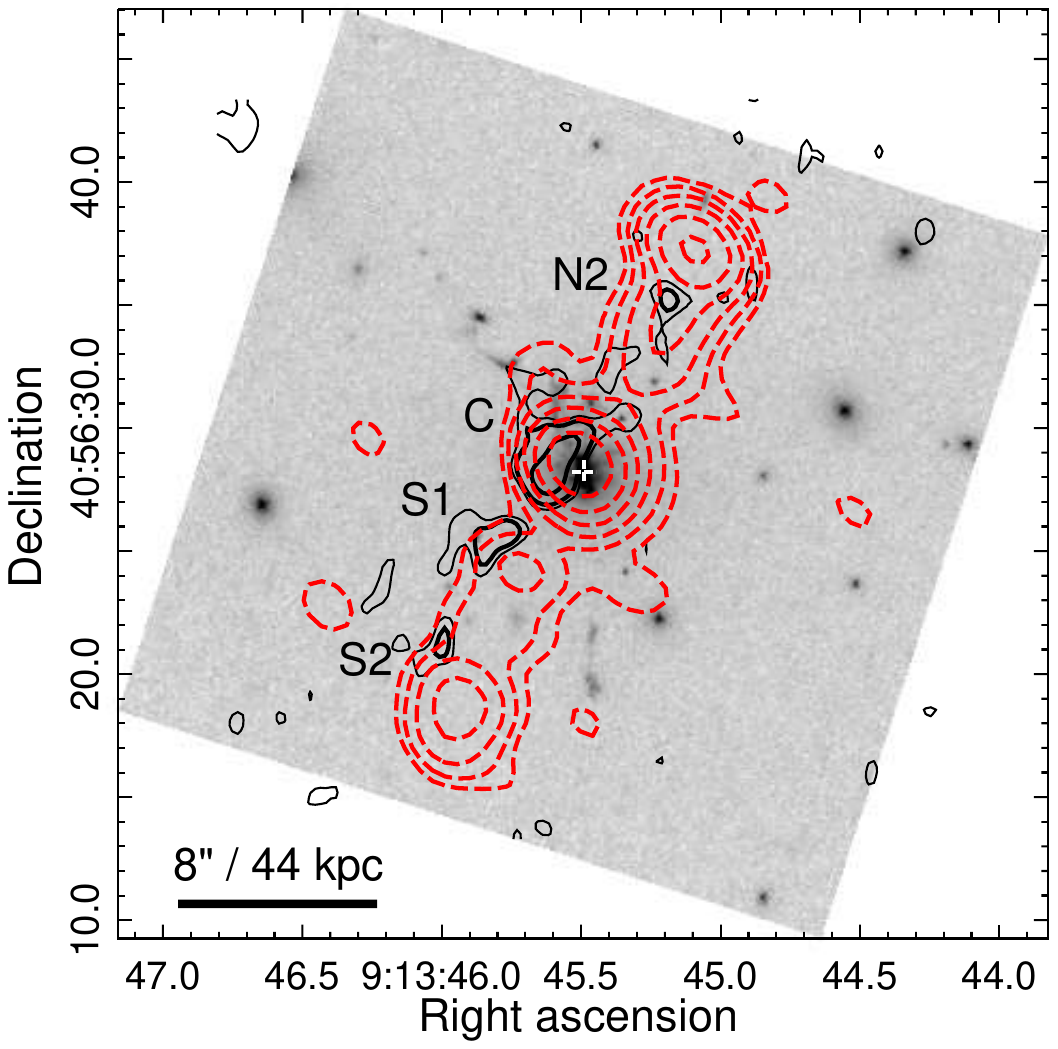}
\caption{\label{fig:opt_rad_CO}\hst\ \textit{F622W} image of \IRAS\, overlaid with NOEMA CO(2-1) contours (black, levels as in Figure~\ref{fig:COflux}) and GMRT 1.28~GHz contours (dashed red, from OS12) starting at 3$\times$rms and increasing in steps of factor 2, with rms=20$\mu$Jy~bm$^{-1}$. The cross indicates the position of the AGN.}
\end{figure}

\begin{figure}
\includegraphics[width=\columnwidth,viewport=60 150 555 645]{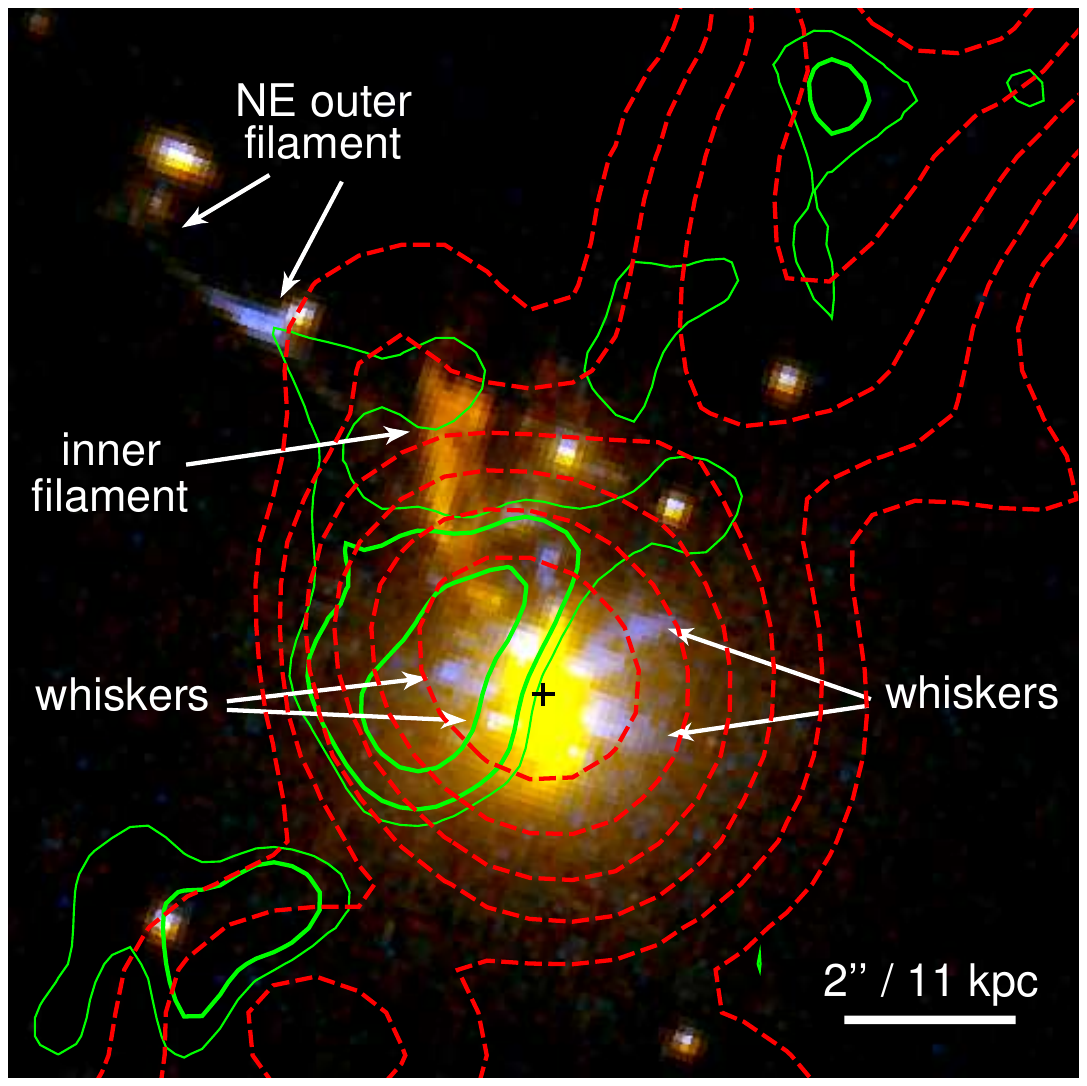}
\caption{\label{fig:zoom}False colour \hst\ PC2 image of \IRAS, zoomed in to show the optical filaments northeast of the BCG. An \textit{814W} filter image is used for the red band, a \textit{622W} image for green and a scaled \textit{622W} subtracted from the \textit{814W} image for the blue band. This combination is chosen to emphasize the optical filaments referred to in the text. NOEMA CO(2-1) (green) and GMRT 1.28~GHz contours (dashed red respectively, from OS12) are overlaid. The cross indicates the position of the AGN.}
\end{figure}

\subsubsection{Comparison with radio continuum, X-ray and optical structures}

Figure~\ref{fig:opt_rad_CO} shows the contours of the low-velocity CO(2-1) emission overlaid on an \hst\ \textit{F622W} image of the cluster core, with GMRT 1.28~GHz radio continuum contours showing emission from the old radio jets and lobes associated with the BCG. Further description of the \hst\ and GMRT data used in this section can be found in OS12. The majority of the CO emission, including clumps C, S1 and S2, falls along the east side of the radio jets. Clump N2 is located at the base of the northern radio lobe, while clump S2 is located on the eastern side of the base of the south radio lobe. It is notable that clump C only overlaps the eastern half of the BCG.

Figure~\ref{fig:zoom} shows the relationship of the BCG and CO clumps in greater detail. Clump C overlaps the base of the brightest of the optical filaments associated with the galaxy (the ``inner filament'') as well as some of the smaller filaments extending out of the BCG (``whiskers''). As mentioned in Section~\ref{sec:intro}, the [O\textsc{iii}] emission from the inner filament has a velocity of $\sim$100\kmps\ relative to the BCG, comparable to the velocity of the molecular gas. It therefore seems plausible that the two are cospatial, though the CO appears more extended than the ionized gas. At the 2$\sigma$ significance level, there is a suggestion of an extension northeast from clump~C, in a similar direction to that of the inner filament and northeast optical filament. This would be interesting if real, but the low significance makes any association speculative. Deeper \hst\ observations and NOEMA observations would perhaps give a clearer view of the connection between the two gas phases.

\begin{figure}
\includegraphics[width=\columnwidth,viewport=60 150 555 645]{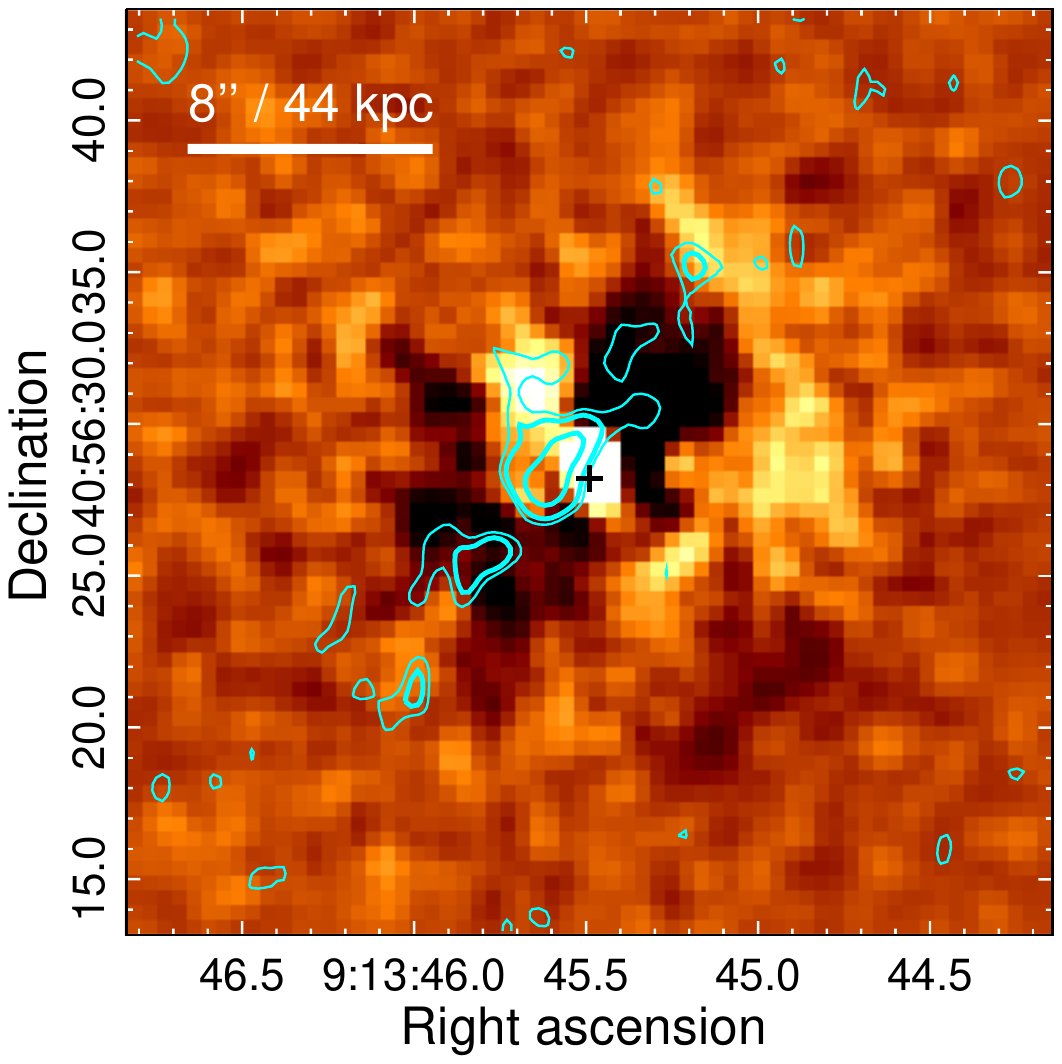}
\caption{\label{fig:Xresid}\chandra\ 0.3-3~keV residual image, smoothed with a 2-pixel ($\sim$1\arcs) Gaussian, with NOEMA CO(2-1) contours (cyan, levels as in Figure~\ref{fig:COflux}) overlaid. The cross indicates the position of the AGN.}
\end{figure}

Figure~\ref{fig:Xresid} shows the CO contours overlaid on a lightly smoothed \chandra\ 0.3-3~keV residual map, after subtraction of the best-fitting elliptical $\beta$-model (see OS12 for details, and their Fig.~7 for a comparison with the GMRT radio contours). The dark regions northwest and southeast of the AGN are interpreted as cavities inflated by the old radio jets. They only extend along the jets and are not correlated with the radio lobes, but as discussed in OS12 the \chandra\ exposure may be insufficient to detect cavities on the scale of the lobes, which are relatively small. Clump S1 falls within the southern cavity and N2 is at the outer edge of the northern cavity. Clump N2 falls on a linear X-ray surface brightness excess structure, while clump C partly overlaps another region of excess X-ray emission extending northeast from the BCG, corresponding to the inner optical filament. The linear structure close to clump N2 could be gas compressed by the expansion of the cavity, or could be associated with sloshing motions. The brighter structure overlapping clump~C seems more likely to be a region of rapid cooling; this would explain the apparent spatial correlation between X-ray, [O\textsc{iii}], and CO emission.

\subsection{Continuum emission}
\label{sec:dust}
\begin{figure}
\includegraphics[width=\columnwidth,bb=55 145 560 650]{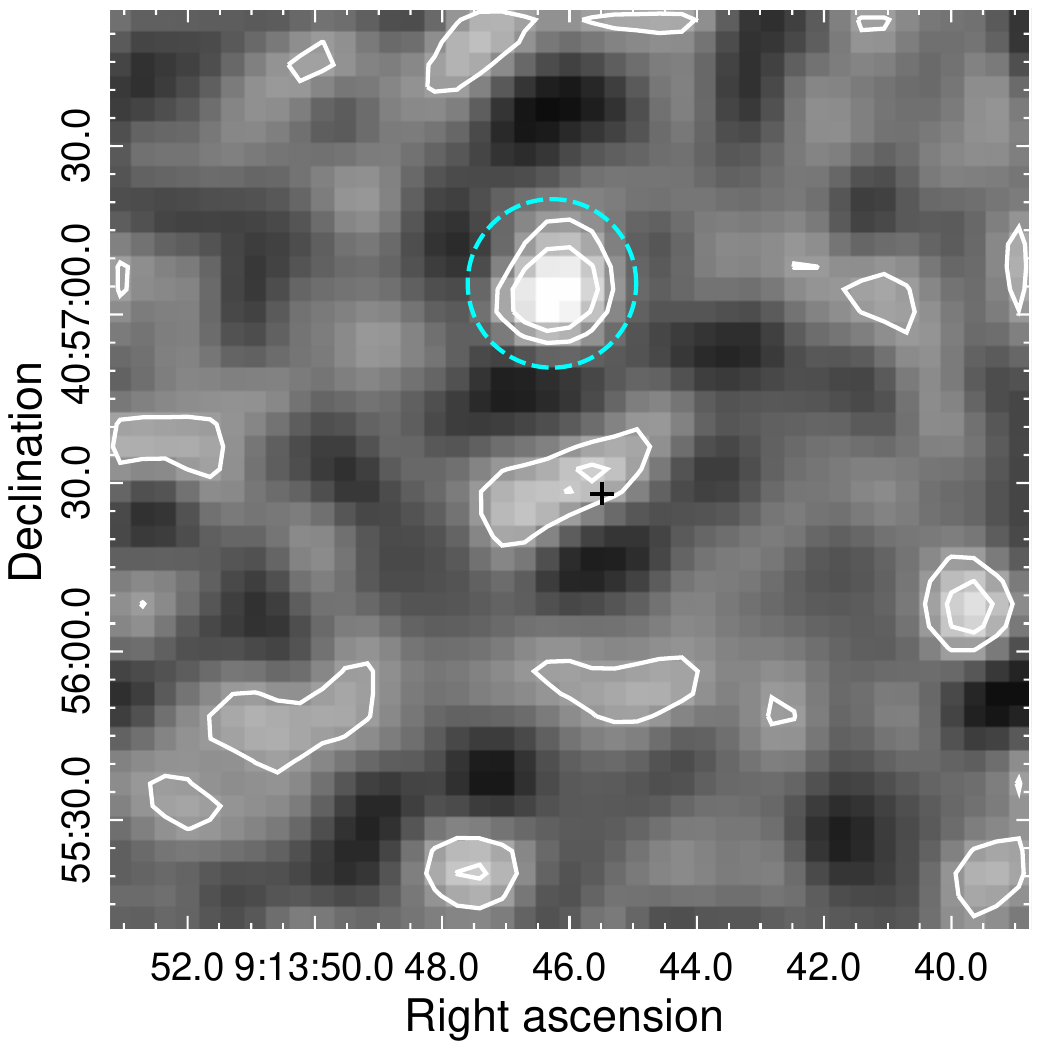}
\vspace{-3mm}
\caption{\label{fig:SCUBA2}JCMT SCUBA-2 850~$\mu$m map of the region around \IRAS, whose position is marked by a cross. Contours levels are 2 and 3$\times$ the noise level, 0.74~mJy~beam$^{-1}$. The circle indicates the effective size of the SCUBA-2 beam (14.6\arcs\ FWHM) and is centred on the nearby source mentioned in the text.}
\end{figure}

Continuum emission is detected from \IRAS\ in both the NOEMA and JCMT SCUBA-2 data, with 159.9~GHz and 850~$\mu$m flux densities of 0.56$\pm$0.03~mJy and 3.355$\pm$0.7~mJy respectively. Figure~\ref{fig:SCUBA2} shows a section of the SCUBA2 map centred on the galaxy. \IRAS\ is among the weaker sources in the 850~$\mu$m map, with a brighter source centred close to WISEA~J091346.94+405703.1, only $\sim$38\arcs\ to the north. The peak of the 850~$\mu$m flux is located slightly to the northeast of the nucleus of \IRAS, but given the size of the SCUBA-2 beam (14.6\arcs\ FWHM), the position is reasonably consistent with that of the galaxy.

\begin{figure*}
\includegraphics[width=\textwidth,bb=36 306 577 487]{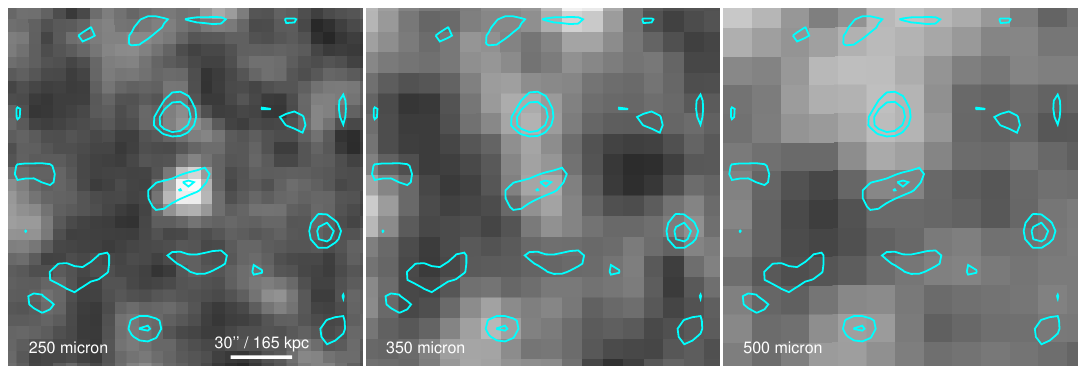}
\caption{\label{fig:Herschel}\textit{Herschel} SPIRE 250, 350 and 500~$\mu$m images centred on \IRAS, with SCUBA-2 850~$\mu$m contours overlaid. Contours levels are 2 and 3$\times$ the noise level, 0.74~mJy~beam$^{-1}$.}
\end{figure*}

We have also reanalysed the Herschel Spectral and Photometric Imaging Receiver (SPIRE) data for the system, and the resulting 250, 350 and 500~$\mu$m images are shown in Figure~\ref{fig:Herschel}. At 250~$\mu$m \IRAS\ is relatively bright, while the 850~$\mu$m source to its north is considerably weaker. However, at longer wavelengths the HyLIRG becomes fainter, and the instrument resolution poorer, so that while at 350~$\mu$m the two sources can still be separated, at 500~$\mu$m only the northern source is visible. Table~\ref{tab:IRflux} lists the fluxes and upper limits measured from the NOEMA, SCUBA-2 and SPIRE.

\begin{table}
\caption{\label{tab:IRflux}Measured far-infrared, sub-millimeter and millimeter-wave flux densities for \IRAS from NOEMA, JCMT SCUBA-2 and \textit{Herschel} SPIRE. The wavelength for the NOEMA measurement corresponds to a frequency of 159.9~GHz.}
\begin{center}
\begin{tabular}{lcc}
\hline
Instrument & Wavelength & Flux density \\
           & ($\mu$m)   & (mJy)\\
\hline
NOEMA   & 1875 & 0.56$\pm$0.03 \\
SCUBA-2 & 850 & 3.2$\pm$0.9 \\
        & 450 & $<$17.4 \\
SPIRE   & 500 & $<$24 \\
        & 350 & 15.7$\pm$5.3 \\
        & 250 & 53.3$\pm$4.1 \\
\end{tabular}
\end{center}
\end{table}

\citet{Farrahetal16} explored the infrared spectrum of \IRAS, describing it with two radiative transfer grid models, the first (dominant at shorter wavelengths) an AGN whose radiation is reprocessed by surrounding dust \citep{EfstathiouRowanRobinson95,Efstathiouetal13} and the second a dusty starburst \citep[][dominant in the far infrared]{Efstathiouetal00}. The combination of the two components modelled the infrared emission well (see their Fig.~1), but at wavelengths longer than 250~$\mu$m only upper limits on the flux were available, and the model was thus unconstrained. In the radio, OS12 found the 151~MHz -- 15~GHz spectrum to be well described by a power law with spectral index\footnote{The radio spectral index $\alpha$ is defined as $S_\nu\propto\nu^{-\alpha}$ where $S_\nu$ is the flux density at frequency $\nu$} $\alpha$=1.25$\pm$0.01. 

\begin{figure}
\includegraphics[width=\columnwidth,bb=10 10 780 700]{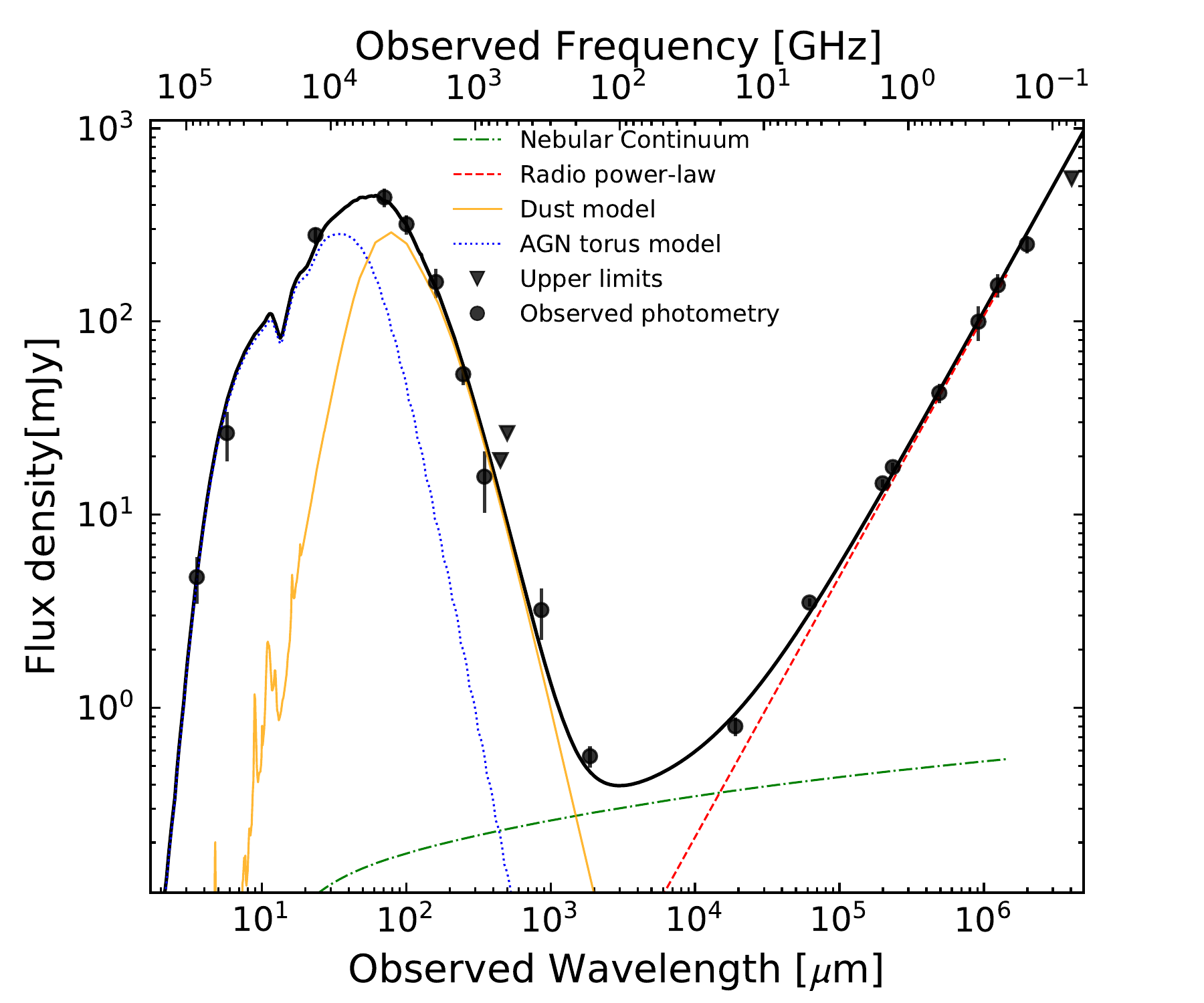}
\caption{\label{fig:radio}Spectrum of the continuum emission from the BCG from the infrared into the radio. Measured fluxes and uncertainties are marked by circles with error bars, inverted triangles indicate upper limits. The thick solid black line shows the fitted model, with the thinner lines indicating the AGN torus (blue dotted), dusty star formation (orange solid), nebular continuum (green dot-dashed), and radio continuum (red dashed) components.}
\end{figure}

Figure~\ref{fig:radio} shows the infrared to radio spectrum of \IRAS, including our continuum flux measurements. Our 159.9~GHz NOEMA and 850~$\mu$m SCUBA-2 continuum measurements clearly belong to the infrared part of the spectrum but comparison with \citet{Farrahetal16} shows that they fall above the flux expected from their starburst model.

We modelled the spectrum using the Code Investigating GALaxy Emission \citep[\textsc{cigale},][]{Boquienetal19}. In addition to the starburst and AGN radiative components used by Farrah et al., we included nebular emission from star-forming regions, and a power law to describe the synchrotron emission from the radio jets. A delayed-$\tau$ star formation history was used with a recent burst (10-50~Myr) accounting for the star-forming population. The \citet{BruzualCharlot03} stellar population library was adopted, assuming a \citet{Chabrier03} initial mass function (IMF). The metallicity was allowed to vary from 10$^{-4}$--5$\times$10$^{-2}$, where 2$\times$10$^{-2}$ corresponds to the solar value. The re-processed dust emission associated with the star formation was modelled using the templates of \citet{Daleetal14}, which are based on a sample of star-forming nearby galaxies, and parameterized by a single parameter $\eta$, defined as dM$_d$(U)$\propto$U$^{-\eta}$dU, where M$_d$ is the dust mass and U is the radiation field intensity. This component successfully fits the cold dust emission as seen in Figure~\ref{fig:radio}.

The SKIRTOR \citep{Stalevskietal16} library for dusty AGN tori, based on 3D Monte Carlo radiative transfer calculations, was used to model the IR emission from the AGN heated dust. The synchrotron emission was modelled by a power law whose slope was allowed to vary in the range 1.0-1.5. The current version of \textsc{cigale} only fits to a maximum wavelength of 1~m, so the two longest wavelength radio fluxes are not included in the fit, but appear reasonably well described by the model anyway. The nebular emission component, representing free-free emission from star forming H$\textsc{ii}$ regions, was modelled using templates based on \citet{Inoue11}, which were generated using models form the \textsc{cloudy} spectral synthesis code \citep{Ferlandetal98}. This component makes a significant contribution at millimetre to centimetre wavelengths, most notably affecting the fit to the 15 and 159~GHz data points. In total, the model has 12 free parameters, and the best fit has $\chi^2$=15.67.

Based on the model fit, the starburst age is 25$\pm$13~Myr, consistent with the upper limit of 50~Myr found by Farrah et al. The best fitting synchrotron powerlaw spectral index is 1.35$\pm$0.06, a little steeper than that found from the radio alone. The total intrinsic infrared luminosity of the AGN radiative component is found to be [4.60$\pm$0.62]$\times$10$^{46}$\ergps, consistent with the value $\sim$4.9$\times$10$^{46}$\ergps\ found by Farrah et al. The half opening angle of the AGN torus model is 40\degree, and an edge-on geometry is favoured, but we note that the combined infrared and X-ray modelling of Farrah et al. is likely to give a more reliable estimate of these geometric parameters than our fit.

\section{Discussion}
\subsection{ICM properties and molecular gas location}
\label{sec:structure}
In general terms, the overall structure of the molecular gas is very similar to that observed in other cool-core galaxy clusters with central FR-I radio galaxies. Typically, a large fraction of the gas is found in or near the BCG, with the remainder in filaments, some of which may wrap around or fall behind the radio lobes and cavities, all at low velocities relative to the galaxy \citep[see e.g.,][]{Olivaresetal19,Russelletal19}. In our case, clump~C contains $\sim$55 per cent of the molecular gas, and all of the clumps (except the high-velocity component) have peak velocities within 100\kmps\ of the BCG. However, the molecular gas distribution around \IRAS\ is significantly more extended than is usual. \citet{Olivaresetal19} find that for 13 clusters with extended molecular gas filaments, the largest extend $\leq$25~kpc. For comparison with another QSO-hosting cluster, the molecular filaments in the Phoenix cluster extend $\sim$20~kpc \citep{Russelletal17}. Clump~C extends $\sim$3\arcs\ (16.5~kpc), while the smaller clumps extend out to $\sim$55~kpc. This is similar to the extent of the H$\alpha$ filamentary nebula complex around NGC~1275 \citep[the BCG of the Perseus cluster, see e.g.,][]{Conseliceetal01,GendronMarsolaisetal18}, many of whose filaments contain molecular gas \citep{Salomeetal06,Salomeetal11}. However, NGC~1275 is the closest example of such a filamentary nebula, and is thus observed in greater depth than any other. The fact that we see a similar extent in \IRAS\ at $z$=0.44 emphasizes the unusual scale of its molecular gas complex.

\begin{figure}
\includegraphics[width=\columnwidth,bb=30 220 560 760]{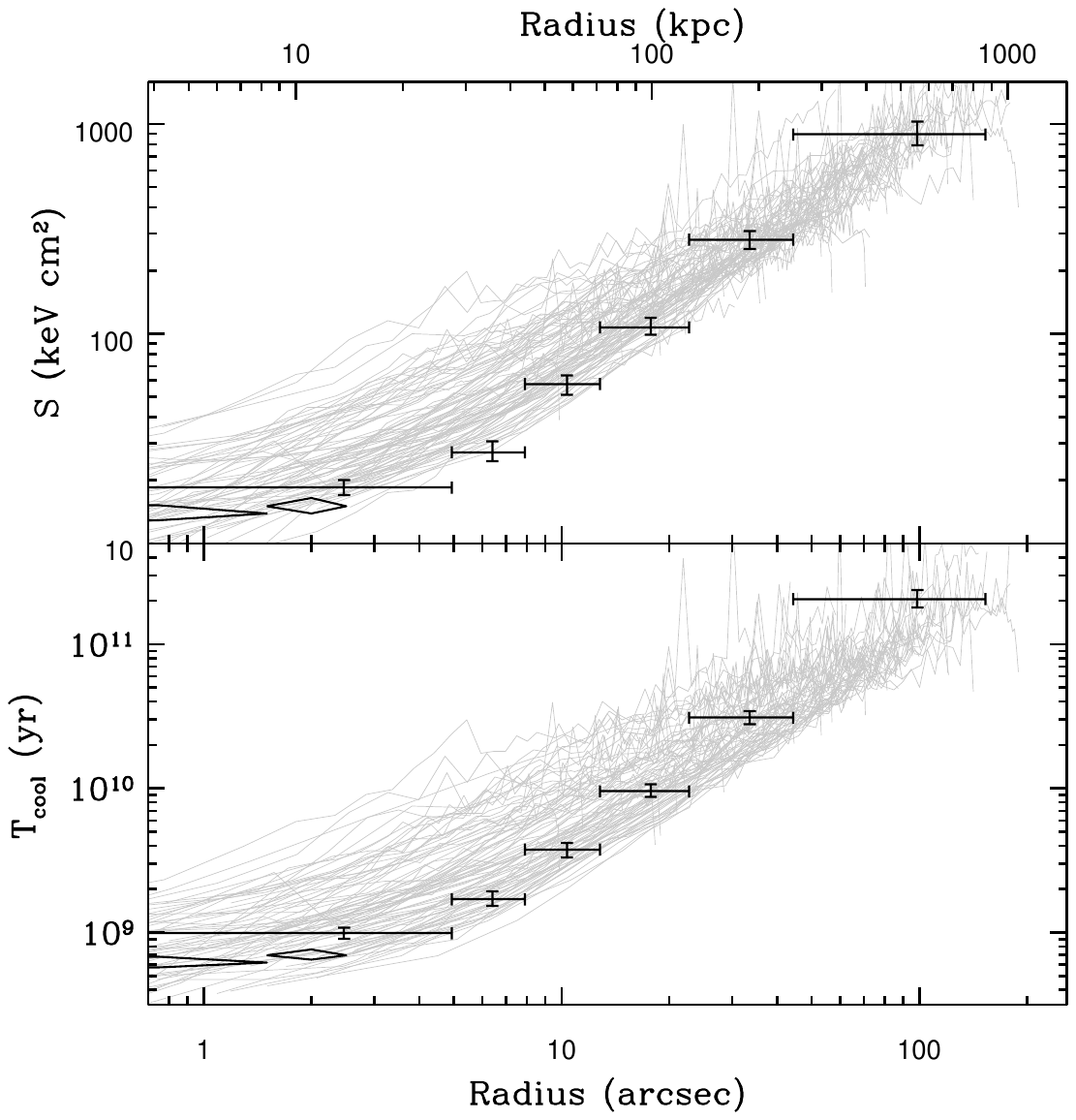}
\caption{\label{fig:STc}Profiles of entropy and isobaric cooling time for the cluster. Crossed error bars show the deprojected profiles from OS12. Grey lines show profiles of low-redshift strong cool core clusters drawn from the ACCEPT sample of \citet{Cavagnoloetal10}. The diamonds show the entropy and cooling time estimated for the 1.5-2.5\arcs\ ($\sim$8-14~kpc) annulus based on the projected temperature and surface brightness profile.}
\end{figure}

The cluster has a strong cool core; OS12 report ICM entropy in the central 5\arcs\ of $\sim$20\kevcmsq, and isobaric cooling time $\sim$1~Gyr. These values are derived from a deprojected spectral profile with limited resolution, but we can estimate values closer to the core, at least roughly. OS12 found a projected temperature in a 1.5-2.5\arcs\ bin, immediately outside the region affected by the QSO, of 3.27$^{+0.31}_{-0.24}$~keV. We fitted a $\beta$-model to the 0.5-7~keV surface brightness profile, excluding a 1.5\arcs\ radius region centred on the AGN and regions covering the jets, lobes and cavities. We find the best fitting model has a core radius of r$_{\rm c}$=4.93$^{+0.12}_{-0.14}$\arcs\ and slope parameter $\beta$=0.593$\pm$0.003. Normalising a 3-dimensional $\beta$-model with these parameters to match the density OS12 found in the central 5\arcs, we estimate the central density to be n$_{\rm e}$$\sim$0.117\pcmcu. Conservatively assuming that the temperature measured in the 1.5-2.5\arcs\ bin also applies within 1.5\arcs, we estimate the entropy in the central 1.5\arcs\ (8.3~kpc) to be $\sim$14~\kevcmsq\ and the isobaric cooling time in this region to be $\sim$620~Myr. In reality, the true value of the temperature, and thus the entropy and cooling time, is likely to be somewhat lower. Figure~\ref{fig:STc} shows these estimates plotted against the deprojected entropy and cooling time profiles from OS12, and profiles for strong cool core (SCC) clusters drawn from the sample of \citet{Cavagnoloetal10}; the cluster appears to have similar properties to the most strongly cooling SCC clusters at low redshift.

By comparison, the Phoenix Cluster has an entropy profile steeper than typical for the ACCEPT sample between $\sim$30-100~kpc, falls below any of the ACCEPT clusters inside $\sim$30-40~kpc, and drops significantly inside $\sim$0.007R$_{200}$ owing to a sharp decline in the inner temperature profile \citep{McDonaldetal15,Prasadetal20}. The entropy profile for \IRAS\ flattens in the centre, but it should be noted that the Phoenix Cluster is a significantly more massive system (M$_{200}$$\simeq$2$\times$10$^{15}$\Msol, compared to 6$\times$10$^{14}$\Msol\ for CL~09104+4109, OS12). Scaled by mass, a similar central entropy dip would only have an extent of $\sim$11.5~kpc ($\sim$2\arcs) in our cluster, too small to be detected with the available \chandra\ data.

\begin{figure}
\includegraphics[width=\columnwidth,bb=30 220 560 725]{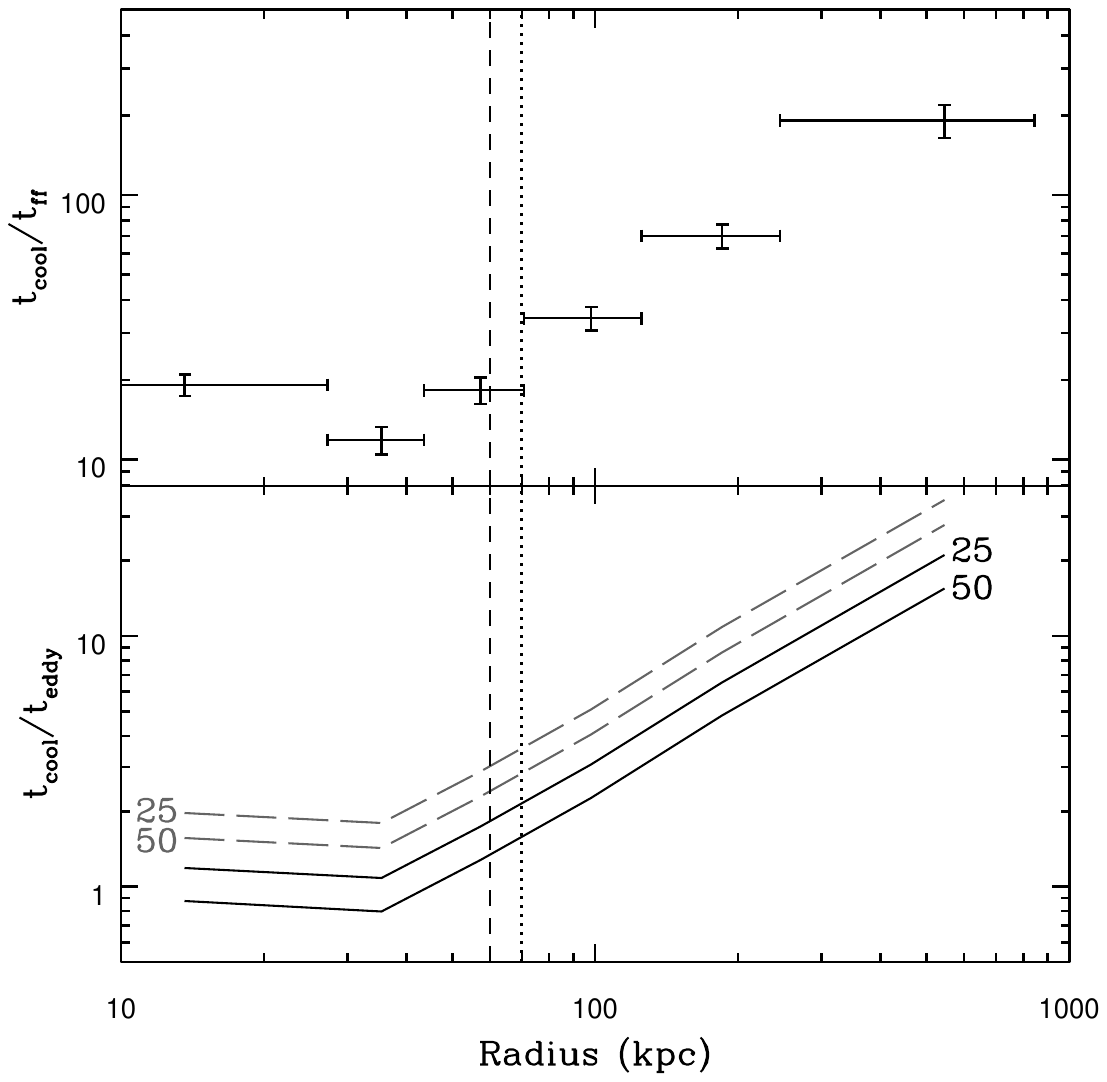}
\caption{\label{fig:tff_teddy}Profiles of the ratio of isochoric cooling time to free fall time (\textit{upper panel}) and eddy time (\textit{lower panel}) in the ICM. t$_{\rm cool}$/t$_{\rm eddy}$ profiles are labelled with the injection length scale ($L$, in kiloparsecs) used in the calculation. Black solid lines are for turbulent velocity dispersion $\sigma_{v,L}$=190\kmps, long-dashed grey lines for $\sigma_{v,L}$=340\kmps. The vertical dashed and dotted lines indicate the radial extent of the CO and old radio jets, respectively.}
\end{figure}

The ratio of the cooling time to the free-fall time has been widely used as an indicator of thermal instability in the ICM, with the extent of filamentary nebulae generally matching the radius at which t$_{\rm cool}$/t$_{\rm ff}$$\la$20 \citep{Voitetal18}. Based on the modelled mass profile of the cluster (see OS12) we can estimate the free-fall time at a given radius as 

\begin{equation}
{\rm t}_{\rm ff} = \sqrt{\frac{2r}{g(r)}},
\end{equation}

\noindent where $g(r)$ is the gravitational acceleration caused by the mass contained within radius $r$. Noting that OS12 calculated the isobaric cooling time, a factor 5/3 longer than the isochoric cooling time that is generally used for comparison with the free fall time, Figure~\ref{fig:tff_teddy} shows the profile of t$_{\rm cool}$/t$_{\rm ff}$ for the cluster. We find that t$_{\rm cool}$/t$_{\rm ff}$$<$20 out to at least $\sim$60 kpc. This agrees well with the extent of the CO clumps. 

\citet{Gasparietal18} argue that the ratio of t$_{\rm cool}$ to the eddy turnover timescale provides a more accurate indicator of thermal instability. As shown by \citet{Prasadetal17}, gas is expected to preferentially condense self-consistently in high density fluctuations caused by the turbulence injected as AGN jets or rising cavities disturb the ICM. Such condensation is expected where the ratio of cooling to eddy timescales approaches unity. The eddy turnover timescale is defined as:

\begin{equation}
t_{\rm eddy} = 2\pi\frac{r^{2/3}L^{1/3}}{\sigma_{v,L}}
\end{equation} 

\noindent where $\sigma_{v,L}$ is the velocity dispersion of the turbulence injected on length scale $L$. Neither of these parameters is directly observable with current instruments. However, $L$ is likely to be comparable to the size of the radio jets or cavities. The turbulent velocities in the condensation regions is likely to drive turbulence in the gas which cools from them, therefore the velocity dispersion of the CO may be a proxy for the ICM. Converting the line-of-sight velocity dispersions measured in clump~C to a three dimensional velocity dispersion suggests that $\sigma_{v,L}$ may be in the range 190-340\kmps. We estimate t$_{\rm eddy}$ for $\sigma_{v,L}$=190 and 340\kmps and $L$=25 or 50~kpc, scales chosen to match the size of the detected cavities or the extent of the CO emission. Figure~\ref{fig:tff_teddy} shows the resulting profiles. We find that t$_{\rm cool}$/t$_{\rm eddy}$ is close to unity throughout the region where CO is observed, supporting the idea that the ICM is thermally unstable and rapidly cooling out to at least 60~kpc.

\subsection{Origin of the molecular gas}
\label{sec:origin}
As mentioned in Section~\ref{sec:intro}, the location at which gas cools out of the ICM and the mechanisms involved in producing molecular and ionized gas filaments in cluster cores are still subjects of debate. It has been suggested that the gas may condense out close to the cluster core and then be uplifted by rising radio lobes, that cool ICM gas is uplifted by the lobes and then condenses out along their line of expansion, or that the jets and lobes disturb thermally unstable gas at larger radii, which condenses \textit{in situ}. For cluster-central quasars there is the additional possibility of ejection of cold gas from the BCG by a radiation-driven outflow.

Considering the last of these options first, many examples of high-velocity molecular outflows from quasars have been observed and although most are on sub-kiloparsec scales \citep[e.g.,][]{Ciconeetal14,Fluetschetal19}, there are examples where molecular clumps have been observed at distances of up to 25-30~kpc and velocities of 1000-1700\kmps\ relative to their AGN \citep{Ciconeetal15,Bischettietal19,Chartasetal20}. Optical spectroscopy of \IRAS\ has shown a $\sim$1000\kmps\ blueshifted component in some of the narrow emission lines of the AGN \citep[e.g.,][]{Kleinmannetal88,CrawfordVanderriest96,Tranetal00}. However, the location of this quasar in the centre of a massive galaxy cluster means that a such an outflow can be effectively ruled out. Any radiation driven wind capable of ejecting molecular clouds out to a radius of 55~kpc would have a dramatic impact on the ICM. The molecular gas would be embedded in much larger plumes of hot gas which would disturb the structure of the ICM and themselves produce X-ray emission. We would expect OS12 to have seen any such plumes in the \chandra\ images and temperature map. As will be discussed later, we also have evidence that the QSO is misaligned with the radio jets, and we believe that the shut down of the jets occurred because the AGN accretion rate increased, causing it to enter its current quasar state. If that is correct, the CO cannot have been transported to its current location by a QSO outflow, since that outflow would have started after the radio jets shut down, and would not be correlated with them.

The distribution of the CO clumps along the old radio jets is generally consistent with scenarios involving uplift of ICM or molecular gas, or with disturbance of thermally unstable gas by the expanding jets. The close correlation of clumps S1 and S2 with the edges of the jets and lobes (see Figure~\ref{fig:opt_rad_CO}) is perhaps most consistent with the disturbance scenario. However, the clumps are not clearly correlated with the X-ray structures, as might be expected if they had formed from the coolest, highest density, or most compressed ICM gas. Clump~N2 is located close to the bright linear filament on the northern boundary of the north cavity, but no clumps are found in the rest of the bright partial rim.

The amount of gas which can be uplifted by a buoyantly rising cavity is in principle equal to the mass of gas the cavity displaces, though simulations suggest that in practice only about half the mass can be lifted. \citet{Popeetal10} describe three mechanisms of uplift: entrainment within rising lobes (probably not relevant in our case), upward drift of gas initially pushed aside by a rising lobe, and wake transport, in which a volume of gas immediately behind the lobe is drawn up with it as it rises, potentially leaving a trail of material along its path. The position of clump N2 comes closest to what we might expect from wake transport.

Estimating the mass of gas displaced by the detected cavities along the jets, and for cavities with the size and position of the radio lobes, we find that the lobes could uplift $\sim$4.3$\times$10$^9$\Msol\ of gas each. This is sufficient to lift clump~N2, but less than the summed mass of clumps S1-2. The detected cavities along the jets are capable of uplifting $\sim$2-5$\times$10$^{10}$\Msol. This would be sufficient to uplift clump S1 and perhaps some fraction of clump C, leaving the cavities in the lobes to uplift clumps S2 and N2. It is therefore at least plausible for the molecular gas to have formed from uplifted material, if there are sizable cavities coincident with the radio lobes. However, it should be noted that the mass of molecular gas we observe would be only a small fraction of the material that would have to be uplifted. For molecular gas to be uplifted it would have to be embedded within, and linked to, a much larger volume of less dense material, whose mass would also have to be lifted. If cool ICM gas was uplifted instead, we would expect only a small fraction to cool out and condense. The plausibility of uplift as an origin for the molecular gas is therefore a question of how much additional low-density material is needed to transport or produce the observed molecular gas.

\textsc{In situ} formation avoids the issues associated with uplift, but has other potential problems, in particular the question of the dust content of the cooling filaments. Filamentary cooling nebulae are observed to contain dust \citep[e.g.,][]{Temietal17,Fogartyetal19}, and dust plays a critical role as a catalyst in the formation of molecular gas \citep[e.g.,][]{GouldSalpeter63,HollenbachSalpeter71}, but unshielded dust grains in the ICM are likely to be rapidly destroyed via sputtering. Dust sources are present in the cluster core, e.g., supernovae and stellar mass loss from stars in the outer halo of the BCG, interstellar medium (ISM) stripped from infalling galaxies. Dust from such sources would need to remain shielded over long periods so as to be available when cooling and condensation was triggered. One possibility is that the combination of shielding by magnetic fields and efficient radiation of heat via molecular line emission could allow packets of dusty cooler gas to survive in the ICM. A comparison can be drawn with the highly multiphase shock region of the galaxy group Stephan's Quintet (HCG~92). In that system, strong collisional shock heating failed to destroy all the molecular gas and dust in an H\textsc{i} filament tidally stripped from a spiral galaxy \citep{Sulenticetal01}, and molecular clouds appear to be able to survive and grow \citep{Guillardetal09} despite being surrounded by a $\sim$0.6~keV thermal plasma \citep{OSullivanetal09}. However, the timescales over which such cool material can survive in the hotter ICM is unclear.

\subsubsection{Lack of velocity gradient}

In principle, we might expect the observed clumps to be made up of collections of smaller, dense molecular clouds which, once formed (or detached from the uplifted gas) would fall under gravity toward the BCG. Their high density and small cross-section would mean that they would not be significantly slowed by the ram-pressure of the surrounding ICM, and would therefore fall ballistically. The free-fall time from their current positions to the BCG nucleus is in the range $\sim$30-90~Myr, while the time required for them to fall from the tips of the lobes to their current locations would be $\sim$50-100~Myr. The high end of these ranges is somewhat smaller than the timescale on which the jets are thought to have shut down, 120-160~Myr, and we might expect molecular gas to have begun forming earlier, or if uplifted to have detached and started to fall back earlier, while the jets were still active.

However, there is evidence that the molecular clouds may be somehow supported by the surrounding medium. \citet{Olivaresetal19} and \citet{Russelletal19} point out that where gradients are seen in molecular filaments they are generally weak, inconsistent with the idea that the gas is in free-fall. Instead they appear more consistent with gas caught in the stream lines behind rising radio lobes, with gas either being drawn outward or beginning to fall back toward the BCG. Previous studies have shown that even where gradients are measured \citep[e.g.,][]{Russelletal16,Russelletal17,Vantyghemetal16,Vantyghemetal18} their inclinations would all appear to be close to the plane of the sky if the filaments were assumed to be in free fall. While any individual system could be aligned so as to present only a small line of sight velocity gradient, this cannot be the case for all clusters. Olivares et al. conclude that the molecular gas infall velocities must be inherently small. It therefore seems likely that the molecular gas retains some connection with its surroundings and is slowed by drag forces, perhaps associated with enveloping layers of warmer ionized gas \citep{Lietal18}, or arising from the magnetic field of the ICM \citep{McCourtetal15}. This would allow for material to be uplifted behind rising lobes, as well as slowing its eventual fall back toward the BCG.

Considering \IRAS\ alone, we are limited in what we can say about the velocity structure of the clouds. The velocities determined from spectral fitting of the clouds along the jets are all consistent within uncertainties. The smoothed velocity map shows only small differences, and on the north side of the BCG we only see clump~N2, so we cannot say whether there might be any gradient between that clump and clump~C. The lack of any gradient may mean that the CO is oriented close to the plane of the sky, but as any gradients are probably inherently small, we cannot place strong limits on orientation. Deeper, higher resolution observations might provide a clearer picture of the velocity structure and the kinematic state of the molecular gas.

\subsubsection{The optical filaments as a locus of cooling}

One difference between \IRAS\ and other systems is that we do not see a clear correlation between the molecular gas and the optical filaments. There is a suggestion of correlation between clump~C and the inner filament, but no indications of optical filaments at the positions of the smaller clumps, or along the radio jets in general. One reason for this is probably the lack of deep H$\alpha$ imaging; owing to the redshift of the cluster, the \hst\ observations used either broad-band filters or a narrow filter chosen to target [O\textsc{iii}], which is significantly weaker line. We may therefore simply be missing a significant fraction of the ionized gas emission. Another possibility is that the detected filaments are illuminated by emission from the AGN, and that other filaments which lie outside the opening angle of its ionisation cones are not intrinsically luminous enough to be detected. Since the radio jets and ionisation cones are not well aligned (as we will discuss in section~\ref{sec:align}) this could explain why we do not observe the expected filaments along the jets. Lastly, it should be noted that if a very extended diffuse CO component is present, and overlaps the detected optical filaments, NOEMA may resolve it out or be insensitive to it. However, we would normally expect the molecular gas to be found preferentially in bright ionized gas filaments, so this does not explain why no filaments are seen at the position of the outer CO clumps.

The position of clump~C on one side of the BCG may indicate that the inner optical filament and bright X-ray excess northwest of the BCG nucleus are sites of rapid cooling from the ICM. \citet{CrawfordVanderriest96} showed that the nebular emission in the inner optical filament has a velocity $<$100\kmps\ offset from the nucleus, and a velocity dispersion of $\sim$250-400\kmps. This agrees fairly well with our modelling of the dominant line component in clump~C ($\langle v\rangle$$\sim$70\kmps, FWHM$\sim$260\kmps). This supports the idea that the two gas phases have a common origin. The negative velocity tail in the spectrum of clump~C suggests there is a component of the gas with a broader velocity dispersion, perhaps indicating a different origin. Alternatively this might represent material which is moving from clump~C deeper into the BCG potential well, gaining velocity as it falls.

\subsubsection{External origin via a gas-rich merger}

Thus far we have discussed the molecular gas in terms of cooling from the ICM. We can also consider the possibility that cool gas has been brought into the BCG via a galaxy merger or interaction. OS12 showed the cluster to be sloshing, probably indicating a recent minor merger or flyby encounter with a group-mass object. They hypothesised that such a merger might have brought a large, gas rich galaxy into the cluster core. However, in a massive cluster such as CL~09104+4109 ram-pressure stripping is quite effective at removing gas from infalling galaxies. Only galaxies on relatively radial orbits are likely to reach the region around the BCG while retaining a significant cool gas reservoir. Galaxies on more circular orbits will take many orbital periods to lose energy through dynamical friction, and their long exposure to ram-pressure seems like to leave them denuded of gas by the time they reach the cluster core. The mass of the cluster (M$_{200}$$\sim$8$\times$10$^{14}$\Msol, OS12) suggests a velocity dispersion for the galaxy population of $\sim$1000\kmps. This is comparable to the sound speed of the ICM in the cluster core, which has a temperature $\leq$4~keV within $\sim$50~kpc. We can expect galaxies on radial orbits to be moving at velocities considerably greater than the cluster velocity dispersion at their closest approach to the BCG, and they would thus be supersonic in the ICM. Examination of populations of jellyfish galaxies in clusters supports these expectations. These galaxies are found to be predominantly recently accreted by the cluster, usually on radial orbits, moving with systematically higher absolute line of sight velocities than the general galaxy population \citep{Jaffeetal18}. Stripping occurs preferentially in the dense ICM of cluster cores, but in these regions the galaxy velocities reach their peak, so the velocity of captured gas relative to the BCG would need to be greatly reduced to bring it down to the low values we see in clump~C. Achieving all of this without destroying the molecular gas through heating and cloud disruption seems unlikely.

\subsection{Orientation of the AGN and jet axis}
\label{sec:align}
The roughly equal length and brightness of the large scale radio jets and lobes suggests that they are aligned relatively close to the plane of the sky \citep{HinesWills93,Hinesetal99} or at least not close to the line of sight. As discussed above, the lack of a velocity gradient in the CO along the jets is not a strong constraint, but it is at least consistent with a jet alignment close to the plane of the sky.

The orientation of the central engine of \IRAS\ has been the subject of much discussion, using detailed multi-wavelength modelling to try to estimate the opening angle and angle with respect to the line of sight of the torus. \citet{HinesWills93} first noted that the old radio jets fall outside the opening angle of the AGN ionisation cones. Based on the corrected polarized flux observed from the QSO ionisation cones, the AGN torus is aligned such that its opening has a projected position angle of 98\degree\ \citep[from west,][]{Hinesetal99}, whereas the axis of the radio jets has P.A.=63\degree. The most recent modelling, combining infrared and X-ray constraints suggests that the AGN is tilted $\sim$35\degree\ toward the line of sight with a half-opening angle also $\sim$35\degree. If the jet axis is in the plane of the sky, the angle between the axis of the jets and the current AGN alignment would therefore be $\sim$45-50\degree. This places the jets outside the current opening angle.

This suggests that as well as changing from a radio galaxy to a QSO over the past 120-160~Myr, the AGN has also undergone a change in its axis. This could be achieved through the merger of a comparable mass supermassive black hole (SMBH). This would imply a major galaxy merger in the recent past of \IRAS. Whether the traces of such a merger would be obvious in the available \hst\ data is unclear. A more likely alternative is that the accretion disk of the QSO was initially misaligned with its rotation axis, causing a realignment of the SMBH. \citet{Babuletal13} discuss the conditions necessary for such a realignment, finding that it is most likely to occur when the accretion rate is high (implying a thin disk and quasar-mode AGN) and SMBH spin relatively low. The SMBH mass estimate of \citet{KongHo18}, M$_{\rm SMBH}$=2.3$^{+10.0}_{-1.9}$$\times$10$^8$\Msol, has large uncertainties, but adopting it and (conservatively) assuming that the SMBH axis has changed by 35\degree, we can estimate the mass of gas which would need to have been accreted via the thin disk \citep[Eqn.~12 of][]{Babuletal13} to be $M_{\rm gas}$=2.3$^{+9.5}_{-1.9}$$\times$10$^6$\Msol. This is a small fraction of the current molecular gas reservoir.

If we assume that the radiative age of the radio jets (120-160~Myr) is a good estimate of the time at which the accretion rate rose and the AGN entered its current QSO state, then the implied accretion rate is only $\sim$0.02\Msolpyr. However, the likely accretion rate over that time is higher. The only evidence of jet activity since the old jets shut down is the 200~pc scale VLBA-detected radio double in the BCG core, so we must assume that the AGN has been a high accretion rate QSO for most of the last 10$^8$~yr. This implies an accretion rate of at least 1 per cent of the Eddington rate, $\sim$0.05\Msolpyr. It is thus likely that the SMBH has accreted at least 6$\times$10$^6$\Msol\ over the past 120~Myr, roughly 2.5 per cent of its current mass, but only 0.01 per cent of the available molecular gas reservoir. Clearly this reservoir is quite sufficient to fuel the QSO, even at much higher accretion rates, for long periods (providing the gas reaches the central engine) and to drive the apparent change in its axis.

\subsection{Star formation}
\citet{Bildfelletal08} found blue colours in the stellar population of \IRAS\ extending out to $\sim$20~kpc. However, this does not mean that the current burst of star formation (SF) extends throughout the galaxy. The blue optical colours likely represent a small mass of young stars formed during the earlier burst of SF which occurred 70-200~Myr ago. Following OS12, we can estimate the size of star forming disk which we would expect given the measured star formation rate (SFR), assuming it follows the standard Kennicutt-Schmitt relation \citep{Kennicutt98},

\begin{equation}
{\rm SFR} = 0.017 \times M_{\rm gas}v_{\rm c}/R
\end{equation}

\noindent where $M_{\rm gas}$ is the mass of gas available to fuel SF, $v_{\rm c}$ is the rotation velocity of the disk, and $R$ is the disk radius. We adopt SFR=110\Msolpyr\ \citep{Farrahetal16}, and $v_{\rm c}$=200\kmps\ based on the [O\textsc{iii}] velocity profile of \citep{CrawfordVanderriest96}. If we consider the whole of clump~C (2.62$\times$10$^{10}$\Msol) as available to fuel SF, this would imply a disk of radius $\sim$830~pc ($\sim$0.15\arcs). More conservatively, if we assume only the negative velocity component of clump~C, which is centred closer to the BCG core, and adopt a value of $\alpha_{\rm CO}$=0.8 (more appropriate for this HyLIRG), the radius may be as low as $\sim$45~pc ($\sim$0.008\arcs). As noted by OS12, a compact, dense, gas-rich star forming disk around the AGN provides a natural explanation for the heavy obscuration of the source seen in X-ray and IR-NUV spectral energy distribution (SED) fitting. \citet{Farrahetal16} argue, based on their combined SED and X-ray spectral modelling, that the vertical height of the obscurer is likely $\sim$20~pc, and its outer edge likely within 125~pc. This is at least comparable to the lower end of the range we estimate for the star forming disk.

An example of a star-forming disk of similar size in a BCG with a filamentary nebula can be found in NGC~1275. Observations of molecular gas have confirmed the presence of a 50-100~pc rotating molecular disk \citep{Nagaietal19,Scharwachteretal13} around the AGN. VLBA observations revealed a roughly circular $\sim$70~kpc radius region of radio continuum emission around the AGN which was initially interpreted as a mini-halo \citep{Silveretal98} but may in fact represent star formation within the molecular disk \citep{NagaiKawakatu21}. \IRAS\ would therefore not be unique if it hosts such a disk.

\subsection{High-velocity CO component}
The high-velocity CO component is only marginally detected, but potentially interesting if its existence can be confirmed. Its origin is unclear. The clumps are $\sim$20~kpc from the AGN, a comparable scale to the largest molecular gas outflows observed around quasars, as discussed in Section~\ref{sec:origin}. However, given what we know about the alignment of the AGN, it seems very unlikely that two clumps, on either side of the AGN, would both have such similar blueshifted velocities; we would expect a velocity gradient across the core. It should also be noted that the two clumps have an alignment similar to the old radio jets, not the ionization cones of the QSO, as would be expected if they had been expelled from the BCG by a radiation-driven wind. The X-ray observations also provide strong evidence against an outflow on this scale. OS12 were able to trace the X-ray surface brightness and temperature profiles on scales smaller than the high-velocity clumps and saw no evidence of such disturbance. An outflow, even on this smaller scale, seems unlikely.

Another possibility is that the molecular gas is the remnant of an infalling galaxy. Molecular line observations of ``jellyfish'' galaxies, cold-gas-rich spirals being ram-pressure stripped as they fall into clusters, have shown that molecular gas can be stripped from the disk, and survive transport several tens of kiloparsecs \citep[e.g.,][]{Jachymetal14,Jachymetal17,Jachymetal19}. We see no clear corresponding galaxy for the gas to have been stripped from, but a galaxy tidally disrupted by a passage close to the BCG might not be obvious in the available data. However, the remaining ionized gas in the stripped galaxy should be more easily detectable. No such components are visible in the \hst\ optical images at or near the location of the CO clumps. Deeper CO observations would be needed to confirm their existence.

\section{Summary and Conclusions}
\IRAS\ is a rare, relatively low-redshift example of a class of system believed to be common at earlier epochs: a QSO at the centre of a cooling flow galaxy cluster. Prior observations have shown that the HyLIRG BCG hosts a type-2 (obscured) QSO, as well as an ongoing, $<$50~Myr old starburst. A 70-200~Myr old stellar population component indicates a previous burst of star formation, while $\sim$70~kpc long radio jets, with a spectral age of 120-160~Myr, suggest that the AGN may have shifted from a radiatively inefficient jet-dominated mode to its current radiatively efficient activity in the recent past. The fuelling of such a system, and its impact on the surrounding cluster, are relevant to the question of how high redshift QSOs can maintain the thermal balance of the ICM. We therefore observed \IRAS\ with the NOEMA interferometer and JCMT SCUBA-2 instrument, to investigate its molecular gas and dust content. Our conclusions are summarized below:

\begin{enumerate}
\item As in many cool core clusters with central FR-I radio galaxies, the CO(2-1) maps show that the molecular gas is primarily located in a series of clumps extending along the path of the rise of the old radio lobes, with a total mass of $\sim$4.5$\times$10$^{10}$\Msol\ (for $\alpha_{\rm CO}$=4.6, or $\sim$7.9$\times$10$^9$\Msol\ for $\alpha_{\rm CO}$=0.8). The velocities of these clumps are within 100\kmps\ of the recession velocity of the BCG, and there is no evidence of any significant velocity gradient along the jets. Roughly 55 per cent of the gas is located in a central clump on the northeast side of the BCG, coincident with the base of the brightest of the optical filaments and with a bright spur in the X-ray residual map. This suggests that this may be a region of rapid cooling from the ICM. The molecular clumps are observed out to $\sim$55~kpc from the nucleus. This is exceptional; molecular gas filaments in low-redshift clusters are not generally observed to extend beyond 25~kpc. The velocity dispersion of the molecular gas clumps is relatively low ($\sim$280-320\kmps) except in the central clump, where a broader tail at low velocities is observed. Comparison with the archival IRAM~30m spectrum suggests that the NOEMA data have captured all the emission from the region covered by the 30m beam, with additional emission coming from clumps at greater distances along the jets.

\item Based on the available profiles of ICM properties from OS12, we find that the molecular gas is contained within a region where the ratio of the azimuthally averaged isochoric cooling time to the free-fall time is t$_{\rm cool}$/t$_{\rm ff}$$\la$25, and the ratio of cooling time to eddy turnover time is likely t$_{\rm cool}$/t$_{\rm eddy}$$\sim$1. We find that cavities associated with the jets and lobes are in principle capable of uplifting a mass of gas comparable to the molecular gas mass we observe. However, when we consider the additional mass of warmer material in which the molecular gas must be embedded, or from which it condensed, this possibility looks less likely. It should also be noted that, while the \chandra\ observation detected relatively large cavities correlated with the jets, it lacked the necessary depth to confirm the presence of cavities associated with the lobes. It seems more plausible that the molecular gas may have formed \textit{in situ}, from thermally unstable gas which was disturbed by the expansion of the jets, provided sufficient dust survives in these regions to catalyse the formation of molecular gas. An unusual feature is the lack of correlation between the molecular and ionized gas, outside the central clump. This may indicate that only the very brightest nebular emission was detected in previous \hst\ observations, or perhaps only those filaments illuminated by the QSO.

\item We find a tentative detection of additional CO(2-1) emission at a relative velocity of $\sim$-1450\kmps, located in clumps $\sim$3\arcs/16.5~kpc northwest and southeast of the nucleus. We briefly consider whether these could be evidence of an outflow driven by the QSO, or a remnant of an infalling, ram-pressure stripped galaxy, but conclude that deeper observations are needed, both to confirm the existence of the clumps and to shed light on their origin.

\item We measure the continuum emission of the BCG at 159.9~GHz with NOEMA and at 850~$\mu$m with SCUBA-2. These two fluxes extend the range of the previously measured far-infrared peak associated with reprocessed emission from the AGN and dusty starburst, but have fluxes well above what would be expected from previous modelling. We model the 3.6~$\mu$m - 1~m spectrum, and find that the observed fluxes likely include contributions from nebular free-free emission arising from the ongoing star-formation in the galaxy. Our modelling suggests a starburst age of 25$\pm$13~Myr, consistent with the previous upper limit of $<$50~Myr. We also find that the spectral index of the powerlaw emission from the radio jets is $\alpha$=1.35$\pm$0.06, somewhat steeper than previously thought.

\end{enumerate}

Based on the X-ray and radio observations of \IRAS, OS12 left open the question of whether its ionized gas filaments were the product of cooling from the ICM or the infall and merger of a gas-rich galaxy, though the latter seemed the less likely option. Our NOEMA observations make clear that ICM cooling has produced a large reservoir of molecular gas in the cluster core, with similar characteristics to those seen in other cool core clusters. The amount of molecular gas in or near the BCG is certainly sufficient to have fuelled the increase in accretion rate which caused the AGN to change from an FR-I radio galaxy to a QSO, and to have caused the realignment of the AGN axis. The extent of the molecular gas is unusually large, and the mismatch between the molecular and ionized gas is surprising. Further observations of the warm ionized component would be useful to investigate this lack of correlation, and determine whether previous observations were simply too shallow. Deeper observations of the molecular gas and hot ICM would also be beneficial. With the former, we could explore whether the various CO clumps are actually linked by lower density material, determine whether the molecular gas distribution in the BCG is really as asymmetric as it appears, and investigate the -1450\kmps\ component. Deeper X-ray observations with \chandra\ have recently been approved, and should allow us to determine the size of any cavities associated with the radio lobes, and thus measure their uplift capacity, as well as investigating the ICM conditions in the bright spur that overlaps clump~C, apparently the most rapidly cooling region in this unique system.

\medskip
\noindent\textbf{Acknowledgments}\\
We thank the anonymous referee for their comments, which materially improved the paper.
E.~O'S. gratefully acknowledges the support for this
work provided by the National Aeronautics and Space Administration (NASA)
through \chandra\ Award Number GO0-21112X issued by the \chandra\ X-ray
Center, which is operated by the Smithsonian Astrophysical Observatory for
and on behalf of the National Aeronautics Space Administration under
contract NAS8-03060.
A.B. acknowledges research support from the Natural Sciences and Engineering Research Council of Canada (NSERC).
This work is based on observations carried out under project number W18DB with the IRAM NOEMA Interferometer. IRAM is supported by INSU/CNRS (France), MPG (Germany) and IGN (Spain).
The James Clerk Maxwell Telescope is operated by the East Asian Observatory on behalf of The National Astronomical Observatory of Japan; Academia Sinica Institute of Astronomy and Astrophysics; the Korea Astronomy and Space Science Institute; Center for Astronomical Mega-Science (as well as the National Key R\&D Program of China with No. 2017YFA0402700). Additional funding support is provided by the Science and Technology Facilities Council of the United Kingdom and participating universities in the United Kingdom and Canada. Additional funds for the construction of SCUBA-2 were provided by the Canada Foundation for Innovation. 

\medskip
\noindent\textbf{Data availability}\\
The NOEMA data used in this work, program W18DB, are available from the IRAM Science Data Archive upon request. The JCMT SCUBA-2 data, programs M18BP056, M15BL114 and M12AC15 are available in the JCMT Science Archive at the Canadian Astronomy Data Centre, \url{http://www.cadc-ccda.hia-iha.nrc-cnrc.gc.ca/en/jcmt/}.

\bibliographystyle{mnras}
\bibliography{../paper}

\clearpage
\appendix
\onecolumn
\section{NOEMA channel maps}
\label{sec:chanmaps}
\begin{figure}
\centering{\includegraphics[width=0.95\textwidth,bb=50 130 542 663]{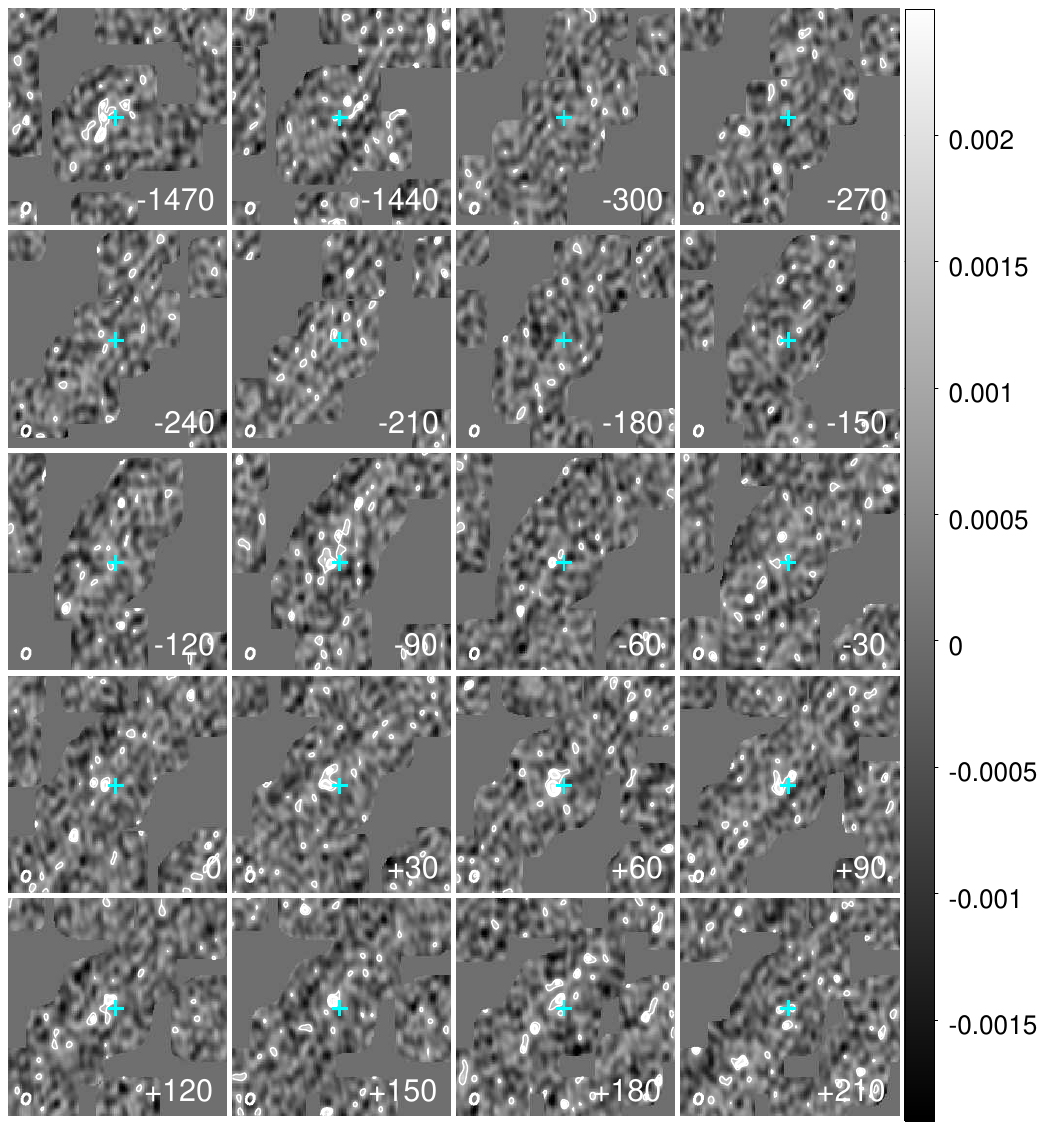}}
\caption{\label{fig:chanmaps} Maps of the 30\kmps\ NOEMA channels used in creating the images in Figure~\ref{fig:COflux}. Solid grey areas are masked as described in Section~\ref{sec:COres}. On each panel, heavy and narrow contours mark the 3$\sigma$ and 2$\sigma$ significant regions. The velocity of each channel in \kmps\ is shown in the bottom right of each panel, the beam size by a dashed ellipse in the bottom left, and the position of the AGN by a cyan cross. All panels share the same orientation and scale as Figure~\ref{fig:COflux}. The colour bar indicates flux density in Jy~km~s$^{-1}$~bm$^{-1}$.}
\end{figure}

\label{lastpage}

\end{document}